\documentclass[structabstract]{aa}  
\usepackage{graphicx}
\usepackage{txfonts}
\usepackage{natbib}
\usepackage{float}
\bibpunct{(}{)}{;}{a}{}{,} % to follow the A&A style

\begin{document}
   \title{New insights into the dust formation of oxygen-rich AGB stars 
\thanks{Based on observations made with Very Large Telescope 
Interferometer (VLTI) at the Paranal Observatory under program IDs 073.D-0711, 
075.D-0097, 077.D-0630, 079.D-0172, and 082.D-0723.}}

   \author{I. Karovicova\inst{1},
          \
          M. Wittkowski\inst{2},
          \
          K. Ohnaka\inst{3},
          \
          D. A. Boboltz\inst{4},
          \
	  E. Fossat\inst{5},
          \and
          M. Scholz\inst{6,7}}

   \institute{Max-Planck-Institut f\"ur Astronomie, K\"onigstuhl 17, 
D-69117 Heidelberg, Germany \email{ikarovic@mpia.de}
         \and 
             European Southern Observatory, Karl-Schwarzschild-Str. 2, 
85748 Garching bei M\"unchen, Germany
         \and 
             Max-Planck-Institut f\"ur Radioastronomie, Auf dem H\"ugel 69, 
53121 Bonn, Germany
         \and 
             United States Naval Observatory, 3450 Massachusetts Avenue, NW,
Washington, DC 20392-5420, USA\\
Present address: National Science Foundation, 4201 Wilson Boulevard, Arlington, VA, 22230, USA
         \and
Laboratoire Lagrange, UMR7293, Universit\'e de Nice Sophia-Antipolis, 
CNRS, Observatoire de la C\^ote d'Azur, 06300 Nice, France
         \and
             Zentrum f\"ur Astronomie der Universit\"at Heidelberg (ZAH), 
Institut f\"ur Theoretische Astrophysik, Albert-Ueberle-Str. 2, 
69120 Heidelberg, Germany
         \and 
             Sydney Institute for Astronomy, School of Physics, 
University of Sydney, Sydney NSW 2006, Australia
             }

   \date{Received 26 July 2013; accepted 30 September 2013}

   \abstract 
%% Context
{AGB stars are one of the major sources of 
dust in the universe. The formation of molecules and dust grains 
and their subsequent expulsion into the interstellar medium via strong 
stellar winds is under intense investigation. This is in particular true 
for oxygen-rich stars, for which the path of dust formation has remained 
unclear.}
%% Aims 
   {
We conducted spatially and spectrally resolved mid-infrared multi-epoch 
interferometric 
observations to investigate the dust formation process in the extended 
atmospheres of oxygen-rich AGB stars.}
%% Methods
   {
We observed the Mira variable AGB stars S~Ori, GX~Mon and R~Cnc between 
February 2006 and March 2009 with the MIDI instrument at the VLT 
interferometer. 
We compared the data to radiative transfer models of the dust shells, 
where the central stellar intensity profiles were described by dust-free 
dynamic model atmospheres.
We used Al$_2$O$_3$ and warm silicate grains,
following earlier studies in the literature.}
%% Results
  {
Our S~Ori and R~Cnc data could be well described by an Al$_2$O$_3$ dust 
shell alone, and our GX~Mon data by a mix of an Al$_2$O$_3$ and a 
silicate shell.
The best-fit parameters for S~Ori and R~Cnc included photospheric angular 
diameters $\Theta_\mathrm{Phot}$ of 9.7$\pm$1.0\,mas and 12.3$\pm$1.0\,mas, 
optical depths $\tau_\mathrm{V}$(Al$_2$O$_3$) of 1.5$\pm$0.5 and 1.35$\pm$0.2,
and inner radii $R_\mathrm{in}$ of 1.9$\pm$0.3\,$R_\mathrm{Phot}$ and 
2.2$\pm$0.3\,$R_\mathrm{Phot}$, respectively.
Best-fit parameters for GX~Mon were $\Theta_\mathrm{Phot}$=8.7$\pm$1.3\,mas, 
$\tau_\mathrm{V}$(Al$_2$O$_3$)=1.9$\pm$0.6, 
$R_\mathrm{in}$(Al$_2$O$_3$)=2.1$\pm$0.3\,$R_\mathrm{Phot}$, 
$\tau_\mathrm{V}$(silicate)=3.2$\pm$0.5, and 
$R_\mathrm{in}$(silicate)=4.6$\pm$0.2\,$R_\mathrm{Phot}$.
Our data did not show evidence of intra-cycle and cycle-to-cycle variability
or of asymmetries
within the error-bars and within the limits of our baseline
and phase coverage.
     }
%% Conclusions
   {Our model fits constrain the chemical composition and the inner boundary
radii of the dust shells, as well as the photospheric angular diameters. 
Our interferometric results are consistent with Al$_2$O$_3$ grains 
condensing close to the stellar surface at about 2 stellar radii, 
co-located with the extended atmosphere and SiO maser emission,
and warm silicate grains at larger distances of about 4--5 stellar radii. 
We verified that the number densities of aluminum can match that of the
best-fit Al$_2$O$_3$ dust shell near the inner dust radius in
sufficiently extended atmospheres,
confirming that Al$_2$O$_3$ grains can be seed particles for 
the further dust condensation.
Together with literature data of the mass-loss rates, our sample 
is consistent with a hypothesis that stars with low mass-loss rates form 
primarily dust that preserves the spectral properties of Al$_2$O$_3$, and 
stars with higher mass-loss rate form dust with properties of warm silicates.
}

   \authorrunning{Karovicova et al.}
   \titlerunning{Dust formation process of oxygen-rich AGB stars} 
   \maketitle

%________________________________________________________________

\section{Introduction}
\label{sec:intro}
Stars on the asymptotic giant branch (AGB) are low and intermediate mass stars 
at the end of their stellar lives, rapidly loosing mass via strong stellar
winds.
These winds from AGB stars are one of the major sources of dust
in the Universe. The mass loss and the related processes are currently
under extensive investigation.
Mira variables are AGB stars which pulsate with long periods around several
hundreds days and large amplitudes up to several magnitudes
\citep{Habing2003}. It is widely accepted that pulsations and dynamic 
effects, including shock waves propagating through the stellar atmosphere, 
lift the upper parts of the atmosphere and create a dense and cool environment
suitable for the formation of molecules and dust grains by
condensation from the gas phase. Due to its large opacity, the dust
absorbs the radiation pressure from the star and drags along the
surrounding gas as it is pushed away \citep{Hofner2009}. 
The mass is expelled via a 
dense and dusty outflow from an extended stellar atmosphere with rates of up to
10$^{-4}$M$_{\odot}$/year \citep{Matsuura2009}, and expansion
velocities of 5-30 km/s \citep{Hofner2005}.
However, the corresponding key molecules as well as the
detailed nucleation mechanisms are different for oxygen-rich and
carbon-rich stars. While carbon grain formation in winds of
carbon-rich stars is rather well defined with various forms of 
carbon or hydrocarbons evolving and finally resulting in macroscopic
carbon grains, the dust formation process in oxygen-rich outflows is
not as clear \citep{Woitke2006,Hofner2007}. Although several 
condensation calculations and observations were recently performed 
in order to explain the observed dust components and to predict the 
presence of as yet unidentified dust species 
\citep{Hofner2008,Goumans2012,Bladh2012,Bladh2013,Sacuto2013,Gail2013}, 
the formation and growth 
of dust grains in oxygen-rich stars is still a widely unsolved problem. 

The spectral energy distributions (SEDs) of oxygen-rich AGB stars are known
to show a diversity of characteristic shapes. Based on IRAS data, 
\citet{Little-Marenin1990} classified AGB stars
into several groups depending on different features of the SED.
\citet{Lorenz-Martins2000} were successful to describe the SEDs of a 
number of oxygen-rich Mira variable AGB stars using Al$_2$O$_3$ dust 
shells, silicate dust shells, or a mix thereof, where the Al$_2$O$_3$ 
dust reproduces
a characteristic broad feature from 9\,$\mu$m to 15\,$\mu$m, and
the silicate dust reproduces characteristic features (the 'silicate
features') near 9.7\,$\mu$m and 18\,$\mu$m. This approach is supported by 
theoretical 
thermodynamic calculations where Al$_2$O$_3$ condenses at relatively
high temperatures of $\sim$ 1400\,K and different kinds of 
silicates at lower temperatures below $\sim$ 1200\,K 
\citep[e.g.,][]{Tielens1998,Gail2010}. 

Interferometry in the mid-infrared domain was shown to be well 
suited to probe the atmospheres and dust shells around AGB stars since it is
sensitive to the chemical composition and geometry of dust shells,
their temperature, inner radii, radial distribution, and the mass-loss
rate \citep[for oxygen-rich stars, e.g.,][]{Danchi1994, Monnier1997, 
Lopez1997, Monnier2000, Tuthill2000, Tevousjan2004, Ohnaka2005, Weiner2006, 
Wishnow2010, Zhao-Geisler2011, Zhao-Geisler2012}.

Using optical and near-infrared interferometric polarimetry, 
\citet{Ireland2005} and \citet{Norris2012} confirmed the presence 
of dust with large grain sizes of $\sim$\,0.3\,$\mu$m at small radii 
below three and below two stellar radii, respectively. These radii correspond
to condensation temperatures above that of the usual warm silicates.
They suggest that their observations may be consistent 
with scattering iron-poor magnesium-rich silicates (forsterite) or 
with Al$_2$O$_3$ dust, both of which are transparent at
wavelengths of 1--4\,$\mu$m. 

\citet{Hofner2008, Bladh2012, Bladh2013} also suggested
micron-sized iron-free silicates as wind-driving grains via scattering, motivated by an 
insufficient radiation pressure of usual silicates to drive the winds of 
oxygen-rich AGB stars \citep{Woitke2006, Hofner2007} and the low 
abundance of aluminum. \citet{Sacuto2013} compared dynamic model atmospheres including
such dust grains to mid-infrared interferometry of the M-type
AGB star RT Vir, and succeeded to model the observed interferometric
visibility data in an ad-hoc manner only by adding a substantial amount of Al$_2$O$_3$ grains.

An alternative scenario was proposed by \citet{Gail2013} based on 
cluster formation of SiO as seed particles for silicate dust together
with revised (lower) calculated condensation temperatures of silicate
dust. \citet{Goumans2012} favor the nucleation of magnesium-rich iron-free 
silicates via heteromolecular condensation based on Mg, SiO, and H$_2$O.

It is also not yet understood whether there is a relation between
dust formation and stellar pulsation, i.e. whether dust formation
occurs preferentially at certain phases, or only during some cycles.
\citet{Lopez1997,Tevousjan2004,Wittkowski2007} reported temporal 
variations of the dust located close to the photosphere for oxygen-rich
sources. Monitoring observations by \citet{Karovicova2011} did not show
an indication of flux and variability variations and confirmed that the
expected variations are below their measurement uncertainties.

Various infrared interferometric observations revealed deviations from 
spherical symmetry already in the AGB phase 
\citep{Lopez1997,Monnier2004,Ragland2006,Wittkowski2011,Paladini2012}.
Non-spherical structures in the molecular and dust shells are theoretically 
predicted by three dimensional radiation hydrodynamic simulations of the 
convective interior and the stellar atmosphere of AGB stars 
\citep{Freytag2008}, as well as by pulsation- and shock-induced weak
chaotic motion in the extended atmosphere 
\citep{Icke1992,Ireland2008,Wittkowski2011}. 
\citet{Ohnaka2012} detected 
asymmetry in the M7 giant BK~Vir in the 2.3\,$\mu$m CO line-forming region using
the AMBER instrument at the VLTI. The asymmetry manifests 
itself as non-zero/non-$\pi$ differential and closure phases.
\citet{Sacuto2013} reported on the deviation from spherical symmetry in the 
atmosphere of the M-type semi-regular variable RT~Vir. Interferometric 
measurements of non-zero differential phases (up to 90~$\degr$) using MIDI 
at the VLTI originated most likely from the presence of one or several clumpy 
structures in the vicinity of the star. 

The main goal of the present study was to investigate the dust formation 
process in the atmospheres of oxygen-rich AGB stars using multi-epoch
spatially and spectrally resolved mid-infrared interferometric observations.
First results from our project were reported by \citet{Wittkowski2007} who 
presented MIDI observations at 4 epochs of the oxygen-rich AGB star S~Ori 
and by \citet{Karovicova2011} 
who presented MIDI observations at 13 epochs of the oxygen-rich AGB star 
RR~Aql. These observations were successfully compared to a radiative transfer 
model of the dust shell, where the central 
stellar intensity profile was described by a series of dust-free dynamic 
model atmospheres based on self-excited pulsation models.
Following the successful comparison to IRAS SEDs by 
\citet{Lorenz-Martins2000}, Al$_2$O$_3$ and warm silicate grains
were used in these studies.
In this paper we present further multi-epoch observations of additional 
phases of S~Ori, and of the oxygen-rich AGB stars GX~Mon and R~Cnc. 
These observations were coordinated with VLBA measurements of SiO
masers toward these sources, providing additional constraints on the
conditions and kinematics within the extended atmosphere where the
dust forms. These VLBA measurements will be described in a forthcoming
paper.
%MWI: Double-check with Dave whether it is fine with him to mention
%the coordinated VLBA observations in this way (sentence above).
Using all results form this project, we discuss the dust condensation 
sequence as a function of the distance from the stellar surface and as a 
function of the mass-loss rate.

\section{Characteristics of S~ORI, GX~Mon, and R~Cnc}
\label{sec:sori}

\paragraph{S~Ori:}
S~Ori is an oxygen-rich Mira variable star with spectral type
M6.5e-M9.5e and $V$ magnitude 7.2--14.0 \citep{Samus2009}. 
We adopted a period of $P\sim$\,430 days and a Julian Date of last 
maximum brightness $T_0=$\,2~453~190\,days derived from the
AAVSO\footnote{http://www.aavso.org} and
AFOEV\footnote{http://cdsweb.u-strasbg.fr/afoev} data for the cycles
of our observations.
\citet{Wittkowski2007} presented the first multi-epoch
study of mid-infrared observations using the MIDI instrument at the
VLTI and concurrent radio interferometric observations using the
VLBA, and reported on a
phase-dependence of photospheric radii and dust shell parameters. They
showed that Al$_{2}$O$_{3}$ dust grains and
SiO maser spots form at relatively small radii of $\sim$1.8--2.4 
photospheric radii.
The results suggested increased mass loss with dust
formation close to the surface near the minimum 
visual phase, when Al$_{2}$O$_{3}$ dust grains were co-located with the 
molecular gas and the SiO maser shells, and a more expanded dust shell 
after visual maximum.
\citet{vanBelle1996}, \citet{Millan-Gabet2005}, and \citet{Boboltz2005} 
measured the
near-infrared $K$-band uniform-disk (UD) angular diameter $\Theta_\mathrm{UD}^K$ of
S~Ori to values between 9.6 mas and 10.5 mas at different phases. The distance
toward S~Ori is not precisely known, and we adopted the distance
estimated by \citet{vanBelle2002} of 480\,pc $\pm$ 120\,pc based on a
calibration of the period-luminosity relationship by
\citet{Feast1989}. \citet{Young1995} estimated a mass loss rate of 
2.2\,$\times$\,10$^{-7}$ M$_{\odot}$/year.

\paragraph{GX~Mon:}
\label{sec:gxmon}
The oxygen-rich Mira variable GX~Mon is not very
well investigated. Due to the relatively faint \textit{V} magnitude,
the light curve of GX~Mon is not monitored, and therefore the visual
phase of GX~Mon is not well known. GX~Mon is a star with spectral type
M9 and \textit{V} magnitude 13.2--17.0. The mean
pulsation period $P$ is uncertain, estimated to 
527\,days \citep{Samus2009}. The
distance toward GX~Mon is also not well known. \citet{Olivier2001}
estimated a distance of 700\,pc based on the period-luminosity
relationship by \citet{Feast1989} and \citet{Justtanont1994} derived a
distance of 740\,pc from modeling the CO $J = 1-0$ and $2-1$ lines. In
our study, we adopt a mean value, and estimate the uncertainty to
25\%, i.e. we use a distance of 720$\pm$185\,pc. Diameter measurements 
of GX~Mon have so far not been reported. 
Based an an empirical calibration by \citet{vanBelle1999}, we 
estimated the angular photospheric diameter of GX~Mon of 
6.0\,mas to 7.5\,mas.
A mass loss rate of 5.4\,$\times$\,10$^{-6}$ M$_{\odot}$/year was estimated by
\citet{Loup1993}. \citet{Justtanont1994} estimated a mass-loss rate
of 7.2\,$\times$\,10$^{-6}$ M$_{\odot}$/year assuming a luminosity of 10$^4$L$_{\odot}$.

\paragraph{R~Cnc:}
R~Cnc is a Mira variable star with spectral type M6e--M9e, a 
$V$ magnitude of 6.1--11.8 and a period of 362 days \citep{Samus2009}. 
\citet{Wittkowski2011} studied the photospheric and molecular layers
of R~Cnc using near-infrared interferometry and derived a photospheric 
angular diameter of 11.8\,$\pm$\,0.7\,mas at phase 0.3.
We adopt their distance of $d=$280\,$\pm$\,28\,pc, which is based
on the period-luminosity relation by \citet{Whitelock2000}.
\citet{Young1995} estimated a mass loss rate of 2\,$\times$\,10$^{-8}$ 
M$_{\odot}$/year.

\onltab{1}{
  \begin{table*}[tbp]
\caption[]{VLTI/MIDI observation of S~Ori.}
\label{observations_sori}
\centering 
\begin{tabular}{lrrrrlrllrrrr}
\cr \hline\hline
\cr Epoch&date&Time&JD&$\Phi$$_\mathrm{vis}$&Config.&$B$& Disp.&BC&$B_\mathrm{p}$ &PA&Seeing&$\tau$$_{0}$\\
&& [UTC]&&&&&Elem.&&[m]&[$\degr$]&[$''$]&[msec]\\
&DD/MM/YYYY& &&&&&&&&&&\\\hline
A & 16/02/2006 & 03:29 & 2453783 & 0.38 & E0-G0 &16 & Prism & HS & 12.77 & 71.93 & 0.93 & 4.8 \\
A & 22/02/2006 & 00:17 & 2453789 & 0.40 & E0-G0 &16 & Prism & HS & 15.98 & 72.45 & 1.61 & 1.5 \\
B & 21/09/2006 & 09:22 & 2454000 & 0.89 & K0-G0 &64 & Prism & HS & 60.94 & 70.26 & 1.00 & 1.1 \\
C & 17/10/2006 & 07:32 & 2454026 & 0.95 & H0-D0 &64 & Prism & SP & 60.34 & 69.90 & 1.02 & 2.4 \\
C &            & 08:43 & 2454026 & 0.95 & H0-D0 &64 & Prism & SP & 63.93 & 72.49 & 0.79 & 2.4 \\
C & 18/10/2006 & 08:58 & 2454027 & 0.95 & E0-G0 &16 & Prism & SP & 16.00 & 72.88 & 0.88 & 2.4 \\
C & 19/10/2006 & 07:05 & 2454028 & 0.95 & E0-G0 &16 & Prism & SP & 14.64 & 68.85 & 8.49 & 2.4 \\
C & 20/10/2006 & 09:20 & 2454029 & 0.95 & D0-G0 &32 & Prism & SP & 31.61 & 73.30 & 0.41 & 4.8 \\
C & 14/11/2006 & 05:44 & 2454054 & 1.01 & D0-G0 &32 & Prism & SP & 30.28 & 70.02 & 0.68 & 2.9 \\
D & 17/12/2006 & 02:21 & 2454087 & 1.09 & H0-G0 &32 & Prism & SP & 25.94 & 64.85 & 1.02 & 2.7 \\
D &            & 04:56 & 2454087 & 1.09 & H0-G0 &32 & Prism & SP & 31.99 & 72.79 & 1.18 & 2.3 \\
D & 19/12/2006 & 01:31 & 2454089 & 1.09 & K0-G0 &64 & Prism & SP & 45.13 & 59.80 & 1.02 & 4.6 \\
D &            & 04:46 & 2454089 & 1.09 & K0-G0 &64 & Prism & SP & 63.97 & 72.74 & 0.55 & 8.4 \\
D & 21/12/2006 & 06:30 & 2454091 & 1.10 & H0-G0 &32 & Prism & SP & 28.59 & 73.10 & 0.63 & 5.9 \\
E & 11/01/2007 & 02:57 & 2454112 & 1.15 & E0-G0 &16 & Prism & SP & 15.96 & 72.30 & 1.41 & 2.5 \\
E &            & 04:49 & 2454112 & 1.15 & E0-G0 &16 & Prism & SP & 14.83 & 73.33 & 1.34 & 3.2 \\
E & 13/01/2007 & 03:45 & 2454114 & 1.15 & E0-G0 &16 & Prism & SP & 15.46 & 74.35 & 0.70 & 5.9 \\
E & 17/01/2007 & 01:36 & 2454118 & 1.16 & E0-G0 &16 & Prism & SP & 15.22 & 70.19 & 1.36 & 1.7 \\
E & 18/01/2007 & 04:55 & 2454119 & 1.16 & H0-G0 &32 & Prism & SP & 27.65 & 72.82 & 1.26 & 1.9 \\
E & 19/01/2007 & 01:58 & 2454120 & 1.17 & H0-G0 &32 & Prism & SP & 31.38 & 71.43 & 0.94 & 2.5 \\
E & 20/01/2007 & 04:48 & 2454121 & 1.17 & H0-D0 &64 & Prism & SP & 55.14 & 72.79 & 0.87 & 4.1 \\
E & 21/01/2007 & 04:19 & 2454122 & 1.17 & H0-D0 &64 & Prism & SP & 58.22 & 73.23 & 0.72 & 5.4 \\
F & 10/02/2007 & 00:54 & 2454142 & 1.22 & H0-D0 &64 & Prism & SP & 63.69 & 72.17 & 0.89 & 3.9 \\
F & 11/02/2007 & 00:35 & 2454143 & 1.22 & E0-G0 &16 & Prism & SP & 15.79 & 71.68 & 1.21 & 3.2 \\
F & 12/02/2007 & 01:08 & 2454144 & 1.22 & H0-G0 &32 & Prism & SP & 31.99 & 72.71 & 1.10 & 3.6 \\
G & 12/03/2007 & 23:44 & 2454172 & 1.29 & E0-G0 &16 & Prism & SP & 15.93 & 73.14 & 1.25 & 2.6 \\
G &            & 02:07 & 2454172 & 1.29 & E0-G0 &16 & Prism & SP & 12.30 & 71.46 & 0.87 & 3.5 \\
G & 14/03/2007 & 00:07 & 2454174 & 1.29 & K0-G0 &64 & Prism & SP & 62.32 & 73.41 & 0.88 & 2.8 \\
H & 02/12/2007 & 06:53 & 2454437 & 1.91 & H0-D0 &64 & Prism & SP & 62.08 & 73.43 & 0.89 & 2.1 \\
H & 10/12/2007 & 02:41 & 2454445 & 1.92 & H0-D0 &64 & Prism & SP & 50.63 & 64.00 & 1.43 & 1.9 \\
I & 29/12/2007 & 02:55 & 2454464 & 1.97 & H0-D0 &64 & Prism & SP & 61.12 & 70.36 & 0.94 & 2.7 \\
I & 10/01/2008 & 01:48 & 2454476 & 2.00 & E0-G0 &16 & Prism & SP & 14.85 & 69.32 & 0.80 & 5.9 \\
I & 11/01/2008 & 01:32 & 2454477 & 2.00 & H0-G0 &32 & Prism & SP & 29.02 & 68.59 & 0.87 & 5.4 \\
I &            & 04:12 & 2454477 & 2.00 & H0-G0 &32 & Prism & SP & 31.16 & 73.42 & 0.90 & 6.2 \\
I & 12/01/2008 & 01:50 & 2454478 & 2.00 & E0-G0 &16 & Prism & SP & 15.07 & 69.84 & 1.30 & 4.7 \\
I &            & 03:49 & 2454478 & 2.00 & E0-G0 &16 & Prism & SP & 15.83 & 73.27 & 1.54 & 4.0 \\
J & 06/03/2008 & 00:31 & 2454532 & 2.13 & H0-D0 &64 & Prism & SP & 62.61 & 73.39 & 0.91 & 3.8 \\
J &            & 01:11 & 2454532 & 2.13 & H0-D0 &64 & Prism & SP & 59.51 & 73.35 & 0.94 & 3.6 \\
J &            & 01:29 & 2454532 & 2.13 & H0-D0 &64 & Prism & SP & 57.62 & 73.15 & 0.88 & 3.7 \\
J & 13/03/2008 & 00:29 & 2454539 & 2.14 & H0-G0 &32 & Prism & SP & 30.41 & 73.43 & 1.00 & 2.1 \\
K & 01/04/2008 & 00:24 & 2454558 & 2.19 & H0-G0 &32 & Prism & SP & 26.24 & 72.28 & 0.85 & 2.2 \\
L & 25/12/2008 & 06:00 & 2454826 & 2.81 & A0-G1 &90 & Prism & SP & 72.97 & 123.81 & 0.84 & 4.7 \\
L &            & 06:21 & 2454826 & 2.81 & A0-G1 &90 & Prism & SP & 68.68 & 126.70 & 0.91 & 4.2 \\
L & 31/12/2008 & 03:34 & 2454832 & 2.83 & H0-E0 &48 & Prism & SP & 47.78 & 72.17 & 0.84 & 5.5 \\
L &            & 04:09 & 2454832 & 2.83 & H0-E0 &48 & Prism & SP & 47.95 & 72.95 & 0.76 & 6.2 \\
M & 26/02/2009 & 01:57 & 2454889 & 2.96 & A0-G1 &90 & Prism & SP & 72.09 & 124.37 & 0.75 & 6.0 \\
M &            & 02:12 & 2454889 & 2.96 & A0-G1 &90 & Prism & SP & 69.08 & 126.42 & 0.85 & 5.4 \\
M & 04/03/2009 & 00:20 & 2454895 & 2.97 & E0-H0 &48 & Prism & SP & 47.62 & 73.22 & 1.28 & 3.3 \\ \hline
\end{tabular}
\tablefoot{Observation log for S~Ori. The table lists the epoch, 
the date, the time, the Julian Date (JD), the visual pulsation phase 
$\Phi$$_\mathrm{vis}$, the baseline configuration, the ground length of the 
configuration, the dispersive element, the beam combiner BC, 
the projected baseline length $B_\mathrm{p}$, the position angle on the 
sky P.A. ($\degr$ east of north), the DIMM seeing (at 500 nm), and the 
coherence time $\tau$$_0$ (at 500 nm).}
\end{table*}
}

\onltab{2}{
  \begin{table*}[tbp]
\caption[]{VLTI/MIDI observation of GX~Mon.}
\label{observations_gxmon}
\centering 
\begin{tabular}{lrrrrlrllrrrr}
\cr \hline\hline
\cr Epoch&date&Time&JD&$\Phi$$_\mathrm{vis}$&Config.&$B$& Disp.&BC&$B_\mathrm{p}$ &PA&Seeing&$\tau$$_0$\\
&& [UTC]&&&&&Elem.&&[m]&[$\degr$]&[$''$]&[msec]\\
&DD/MM/YYYY& &&&&&&&&&&\\\hline
A & 16/02/2006 & 04:39 & 2453783 &  & E0-G0 &16 & Prism & SP & 13.99 & 64.87 & 0.70 & 6.6 \\
B & 18/03/2006 & 01:30 & 2453813 &  & E0-G0 &16 & Prism & HS & 15.60 & 70.07 & 1.15 & 5.2 \\
C & 18/10/2006 & 08:14 & 2454027 &  & E0-G0 &16 & Prism & SP & 13.51 & 74.92 & 0.88 & 2.4 \\
C & 11/11/2006 & 07:59 & 2454051 &  & H0-D0 &64 & Prism & SP & 62.26 & 74.22 & 0.70 & 4.0 \\
D & 14/12/2006 & 03:56 & 2454084 &  & E0-G0 &16 & Prism & SP & 12.15 & 74.51 & 1.10 & 1.6 \\
D &            & 06:25 & 2454084 &  & E0-G0 &16 & Prism & SP & 15.95 & 73.30 & 0.91 & 1.9 \\
D & 16/12/2006 & 04:51 & 2454086 &  & H0-G0 &32 & Prism & SP & 28.84 & 74.90 & 1.08 & 2.4 \\
D & 17/12/2006 & 07:11 & 2454087 &  & H0-G0 &32 & Prism & SP & 31.62 & 71.03 & 1.05 & 2.6 \\
D & 19/12/2006 & 03:10 & 2454089 &  & K0-G0 &64 & Prism & SP & 43.80 & 73.80 & 0.58 & 7.9 \\
D & 20/12/2006 & 05:45 & 2454090 &  & K0-G0 &64 & Prism & SP & 63.21 & 73.77 & 0.92 & 3.6 \\
E & 11/01/2007 & 03:32 & 2454112 &  & E0-G0 &16 & Prism & SP & 15.05 & 74.68 & 1.14 & 3.3 \\
E &            & 05:23 & 2454112 &  & E0-G0 &16 & Prism & SP & 15.90 & 71.48 & 0.99 & 4.1 \\
E & 13/01/2007 & 03:45 & 2454114 &  & E0-G0 &16 & Prism & SP & 15.46 & 74.35 & 0.70 & 5.9 \\
E & 18/01/2007 & 03:34 & 2454119 &  & H0-G0 &32 & Prism & SP & 31.18 & 74.18 & 0.93 & 2.6 \\
E &            & 05:14 & 2454119 &  & H0-G0 &32 & Prism & SP & 31.44 & 70.61 & 1.02 & 2.3 \\
E & 19/01/2007 & 02:31 & 2454120 &  & H0-G0 &32 & Prism & SP & 28.49 & 74.94 & 0.66 & 3.5 \\
E &            & 04:53 & 2454120 &  & H0-G0 &32 & Prism & SP & 31.77 & 71.44 & 1.39 & 1.7 \\
E & 21/01/2007 & 02:46 & 2454122 &  & H0-G0 &32 & Prism & SP & 28.22 & 74.95 & 1.04 & 2.4 \\
F &            & 04:18 & 2454141 &  & H0-G0 &32 & Prism & SP & 30.47 & 68.83 & 0.69 & 3.8 \\
F & 10/02/2007 & 01:16 & 2454142 &  & H0-D0 &64 & Prism & SP & 58.35 & 74.87 & 0.78 & 4.4 \\
F &            & 03:09 & 2454142 &  & H0-D0 &64 & Prism & SP & 63.93 & 72.19 & 1.18 & 3.1 \\
F & 11/02/2007 & 00:54 & 2454143 &  & E0-G0 &16 & Prism & SP & 14.06 & 74.93 & 0.77 & 5.1 \\
F &            & 02:36 & 2454143 &  & E0-G0 &16 & Prism & SP & 15.96 & 73.21 & 0.96 & 6.6 \\
F & 12/02/2007 & 01:27 & 2454144 &  & H0-G0 &32 & Prism & SP & 30.10 & 74.68 & 0.79 & 4.6 \\
G & 10/12/2007 & 03:50 & 2454445 &  & H0-D0 &64 & Prism & SP & 44.49 & 73.93 & 1.92 & 1.4 \\
H & 29/12/2007 & 03:57 & 2454464 &  & H0-D0 &64 & Prism & SP & 57.28 & 74.92 & 0.78 & 3.2 \\
H & 10/01/2008 & 02:52 & 2454476 &  & E0-G0 &16 & Prism & SP & 13.78 & 74.93 & 0.92 & 4.8 \\
H & 11/01/2008 & 02:24 & 2454477 &  & H0-G0 &32 & Prism & SP & 25.72 & 74.80 & 0.59 & 7.7 \\
H & 12/01/2008 & 02:40 & 2454478 &  & E0-G0 &16 & Prism & SP & 13.62 & 74.92 & 0.80 & 7.7 \\
H & 13/01/2008 & 07:10 & 2454479 &  & H0-G0 &32 & Prism & SP & 26.45 & 80.27 & 0.60 & 8.3 \\
I & 22/02/2008 & 02:46 & 2454519 &  & H0-D0 &64 & Prism & SP & 63.35 & 71.14 & 1.29 & 2.2 \\
J & 06/03/2008 & 02:02 & 2454532 &  & H0-D0 &64 & Prism & SP & 63.09 & 70.83 & 0.87 & 3.8 \\
J &            & 02:39 & 2454532 &  & H0-D0 &64 & Prism & SP & 60.75 & 68.66 & 1.09 & 3.0 \\
J & 13/03/2008 & 02:32 & 2454539 &  & H0-G0 &32 & Prism & SP & 29.45 & 67.21 & 0.96 & 5.3 \\
J & 14/03/2008 & 02:50 & 2454540 &  & E0-G0 &16 & Prism & SP & 14.12 & 65.29 & 0.63 & 7.3 \\
K & 28/03/2008 & 01:16 & 2454554 &  & E0-G0 &16 & Prism & SP & 15.12 & 68.41 & 1.16 & 2.3 \\
K & 01/04/2008 & 00:43 & 2454558 &  & H0-G0 &32 & Prism & SP & 30.87 & 69.51 & 0.69 & 2.8 \\
K &            & 01:26 & 2454558 &  & H0-G0 &32 & Prism & SP & 28.96 & 66.45 & 0.92 & 2.1 \\
K &            & 02:03 & 2454558 &  & H0-G0 &32 & Prism & SP & 26.71 & 62.90 & 0.82 & 2.3 \\ \hline
\end{tabular}
\tablefoot{The phases are uncertain, and therefore they are omitted.}
\end{table*}
}

\onltab{3}{
  \begin{table*}[tbp]
\caption[]{VLTI/MIDI observation of R~Cnc.}
\label{observations_rcnc}
\centering 
\begin{tabular}{lrrrrlrllrrrr}
\cr \hline\hline
\cr Epoch&DDMMYYYY&Time&JD&$\Phi$$_\mathrm{vis}$&Config.&$B$& Disp.&BC&$B_\mathrm{p}$ &PA&Seeing&$\tau$$_0$\\
&& [UTC]&&&&&Elem.&&[m]&[$\degr$]&[$''$]&[msec]\\\hline
A & 25122008 & 07:04 & 2454826 & 0.94 & A0-G1 & 90& Prism & HS & 84.87 & 110.94 & 0.96 & 3.9 \\
A &          & 08:18 & 2454826 & 0.94 & A0-G1 & 90& Prism & HS & 71.91 & 110.72 & 1.76 & 2.1 \\
A & 31122008 & 04:46 & 2454832 & 0.96 & H0-E0 & 48& Prism & HS & 40.29 & 76.96 & 0.99 & 4.9 \\
A &          & 05:59 & 2454832 & 0.96 & H0-E0 & 48& Prism & HS & 46.27 & 75.16 & 0.70 & 7.3 \\
B & 26022009 & 02:48 & 2454889 & 1.10 & A0-G1 & 90& Prism & HS & 85.91 & 111.11 & 0.53 & 8.7 \\
B &          & 02:58 & 2454889 & 1.10 & A0-G1 & 90& Prism & HS & 84.64 & 110.90 & 0.62 & 7.4 \\
B &          & 04:14 & 2454889 & 1.10 & A0-G1 & 90& Prism & HS & 71.28 & 110.77 & 0.77 & 6.3 \\
B & 03032009 & 02:05 & 2454894 & 1.11 & E0-H0 & 48& Prism & HS & 46.74 & 74.83 & 1.11 & 3.2 \\
B & 04032009 & 00:37 & 2454895 & 1.12 & E0-H0 & 48& Prism & HS & 40.15 & 76.99 & 0.81 & 5.2 \\\hline
\end{tabular}
%}
\end{table*}
}

\section{VLTI/MIDI Observations and data reduction}
\label{sec:observations}

We obtained a total of 97 spectrally-dispersed mid-infrared
interferometric observations, 48 for S~Ori, 40 for GX~Mon, and 9 for R~Cnc. The
observations were conducted using the mid-infrared interferometric
instrument MIDI \citep{Leinert2004}. MIDI combines the beams from two
telescopes of the VLTI \citep{Glindemann2003} and provides spectrally
resolved visibilities in the \textit{N}-band (8-13\,$\mu$m). 
We used the PRISM as a dispersive element with a spectral resolution 
$R = \lambda$/$\Delta\lambda \sim 30$. For most observations, the beams
were combined in \textit{Sci$\_$Phot} (SP) mode, complemented by
several observations in the \textit{High$\_$Sens} (HS) mode. In the
\textit{Sci$\_$Phot} mode beam splitters record the interferograms and
the photometric spectra simultaneously, while in the
\textit{High$\_$Sens} mode the photometric signal is observed after
the interferometric signal.

\begin{figure}
  \includegraphics[height=0.37\textheight,angle=90]{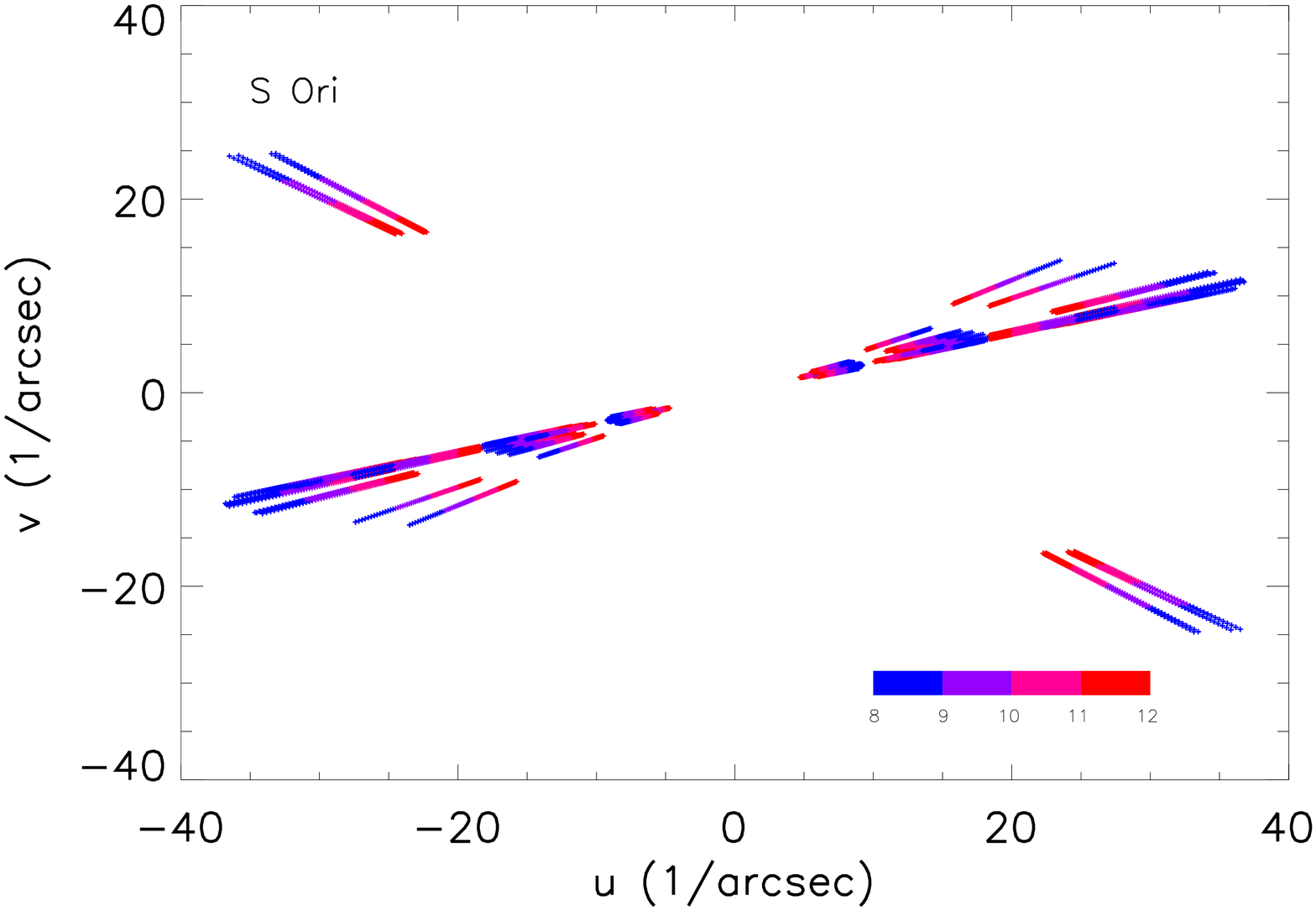}
  \includegraphics[height=0.37\textheight,angle=90]{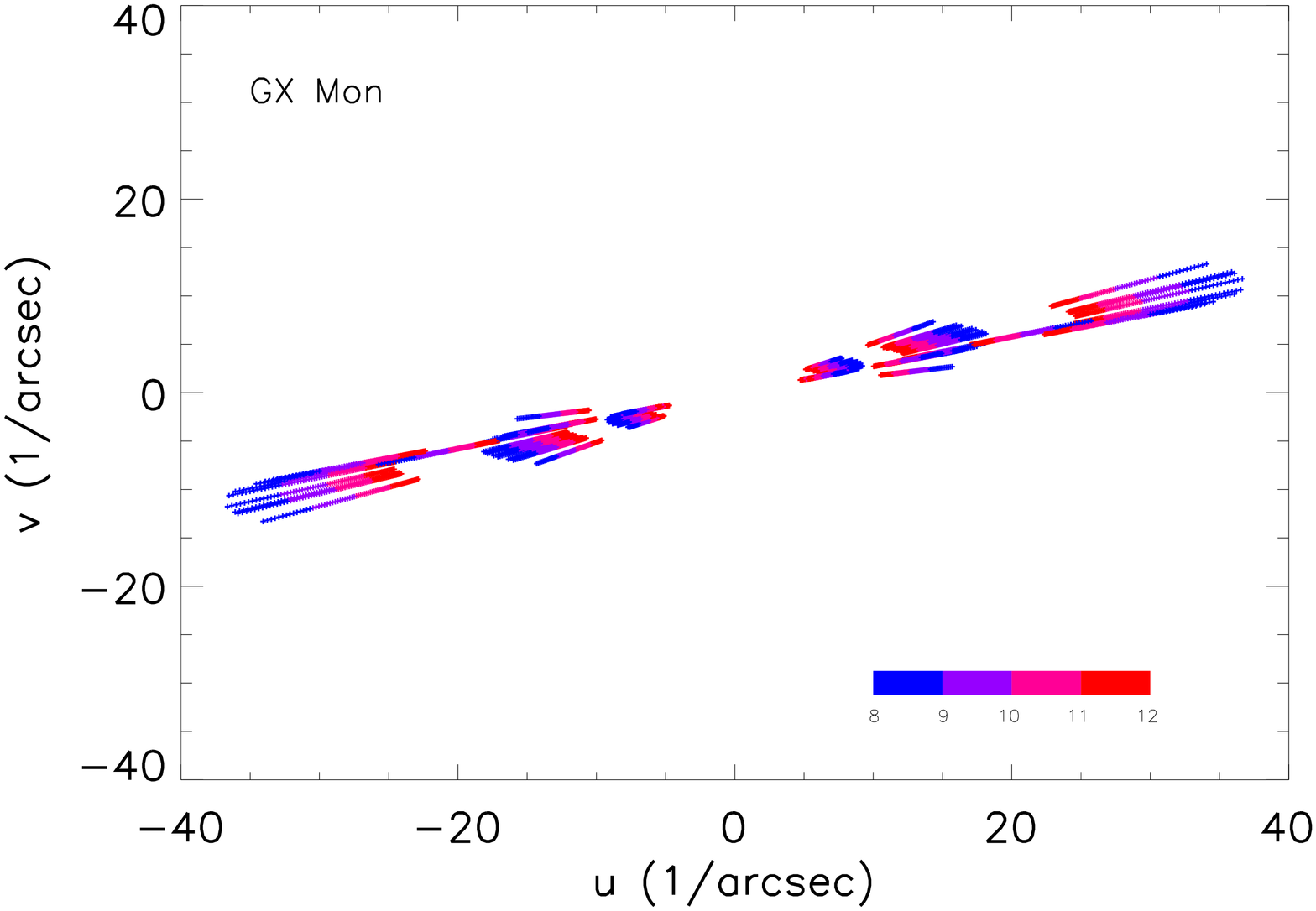}
  \includegraphics[height=0.37\textheight,angle=90]{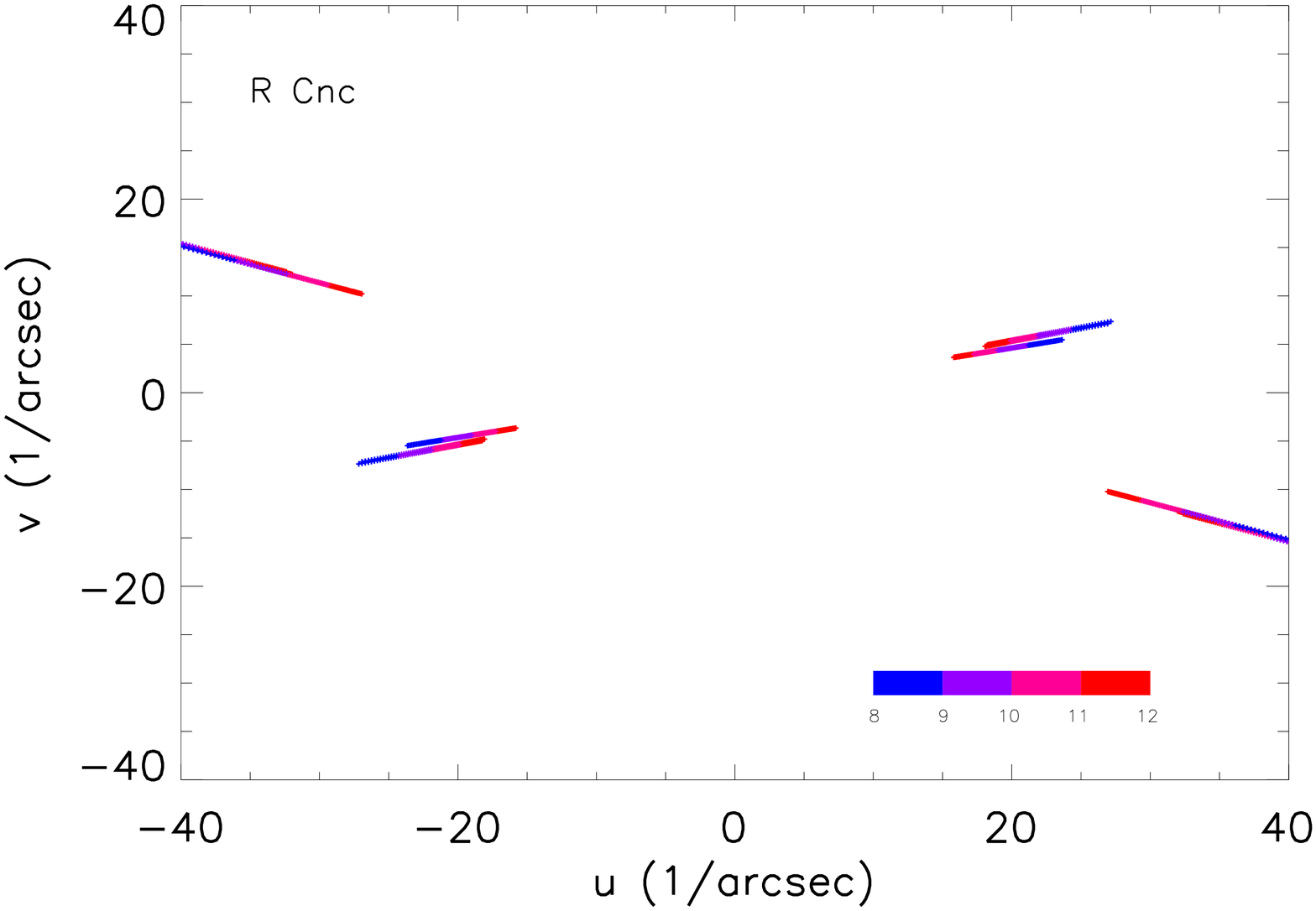}
  \caption[\textit{uv} coverage of S~Ori, GX~Mon, and R~Cnc
  observations]{Coverage of the \textit{uv}-plane of our MIDI
    observations of S~Ori, GX~Mon, and R~Cnc. 
    Each line represents one AT
    observation over the wavelength range from 8~$\mu$m to 13~$\mu$m. 
    The color
    of each point corresponds to the wavelength of the observations,
    as indicated by the color bar.}
         \label{uv_gxmon}
\end{figure}

The details of the observations and the instrumental settings are
summarized in Tables~\ref{observations_sori}, ~\ref{observations_gxmon}, 
and ~\ref{observations_rcnc}. The Tables list the epoch, the date,
the time, the Julian Date (JD), the visual pulsation phase
$\Phi$$_\mathrm{vis}$, the baseline configuration, the ground length of the
configuration, the dispersive element, the beam combiner BC, the
projected baseline length $B_\mathrm{p}$, the position angle on the sky PA
($\degr$ east of north), the DIMM seeing (at 500 nm), and the coherence
time $\tau$$_0$ (at 500 nm). All observations were executed
in service mode and employed different configurations 
of the Auxiliary Telescopes (ATs, 1.8\,m diameter). 
Figure~\ref{uv_gxmon} shows
the \textit{uv} coverage of our new S~Ori, GX~Mon, and R~Cnc observations.

The new observations of S~Ori were obtained between Feb 16, 2006 and Mar 14,
2009 and represent a follow-up of 7 MIDI observations
obtained between Dec 31, 2004 and Dec 30, 2005 by \citet{Wittkowski2007}.
The new observations covered 4 pulsation cycles and were merged
into 14 epochs with a maximum time-lag between individual observations of 
8 days for each epoch (1.9$\%$ of the pulsation period).
For technical problems we had to eliminate 6 observations.
Our new observations of S~Ori cover half a pulsation cycle
from pre-maximum (0.9) to pre-minimum (1.4) phases, but where most
observations were obtained at near-maximum phases of 0.9--1.2.
The four previous S~Ori observations by \citet{Wittkowski2007} covered 
near-minimum and post-maximum phases 0.42, 0.55, 1.16, and 1.27.

The GX~Mon observations were obtained between Feb 16, 2006 and Apr 1, 2008 and 
were merged into 12 epochs with a maximum time-lag between individual 
observations of 11 days ($\sim$2$\%$ of the pulsation period).
We had to eliminate 5 observations for technical reasons. We did not assign 
variability phases to these observations, because the light-curve is not well 
known.

The observations of R~Cnc were obtained at 2 epochs between Dec 25 and Dec
31, 2008 and between Feb 26 and Mar 4 2009, corresponding to near-maximum
phases of 0.95 and 1.10.

For the data reduction we used the MIA+EWS software package, version
1.6\footnote{http://www.strw.leidenuniv.nl/$\sim$nevec/MIDI}. This
package includes two different methods, an incoherent method (MPIA
software package MIA) that analyzes the power spectrum of the observed
fringe signal and a coherent integration method (EWS), which first
compensates for optical path differences, including both instrumental
and atmospheric delays in each scan, and then coherently adds the
fringes. We applied both methods to independently verify the data
reduction results. The detector masks were calculated by the procedure
of MIA, and were used for both the MIA and the EWS analysis.  The
obtained data reduction values correspond to each other, and we chose
to use the results derived from the EWS analysis, which offer error
estimations.

In order to estimate the instrumental visibility losses and to
determine the absolute flux values, we observed immediately before or
after the science target calibrator stars close on sky and with known
diameter and absolute flux values. Our main calibrators were Sirius 
(S~Ori, GX~Mon), HD~81797 (S~Ori, GX~Mon, R~Cnc), and HD~25025 (S~Ori, GX~Mon). 
We adopted a uniform-disk diameter of $\Theta$$_\mathrm{UD}$=6.09$\pm0.03$ 
and a 12 $\mu$m IRAS flux \citep{Beichman1988} of 193 Jy for Sirius, 
$\Theta$$_\mathrm{UD}$=9.14$\pm0.05$ and 158 Jy
for HD~81797, and $\Theta$$_\mathrm{UD}$=8.74$\pm0.09$ and 109 Jy 
for HD~25025. The angular diameters are from the calibrator
list of the MIDI instrument 
consortium\footnote{http://www.ster.kuleuven.ac.be/$\sim$tijl/MIDI\_calibration/mcc.txt}.

Calibrated science target
visibility spectra were calculated using the instrumental transfer
function derived from all calibrator data sets taken during the same
night with the same baseline and instrumental mode as our scientific
target. The number of available transfer function measurements is
related to the number of calibrator stars observed per specific night
including those calibrators observed by other
programs. 
The standard deviation of all transfer function measurements per night 
was used as the uncertainty of the transfer functions. When only one 
calibrator was available, we estimated the uncertainty of the transfer function 
based on typical values when many calibrator stars were observed. The final
errors on the observed visibilities are mostly systematic, and include
the error of the coherence factor of the science target and the
calibrators, the adopted diameter errors, and the standard deviation
of the transfer function over the night.

For most calibrator stars, absolutely calibrated spectra are available
in \citet{Cohen1999}. If the absolutely calibrated spectrum was not
directly available, we instead used a spectrum of a calibrator with a
similar spectral type and similar effective temperature.
The spectra of such calibrators were scaled with the IRAS flux at
12\,$\mu$m to the level of our calibrator. In a few cases, when
the atmospheric absorption was strongly affecting the spectra around
9.5\,$\mu$m, we used another similar calibrator instead of the main
calibrator observed in the same night with the same level of flux.
The ambient conditions for all the observations were monitored, and in
case of problems with clouds, constraints due to wind, significant
differences in seeing, humidity, coherence time, and airmass between
the science target and the corresponding calibrator, the photometry
was omitted from the analysis.

\section{Modeling of the MIDI data}
\label{sec:modeling}

The data were modeled using our established approach,
which was successful to describe the shape and features of the 
flux and visibility spectra of the oxygen-rich Mira variables
S~Ori \citep{Wittkowski2007} and RR~Aql \citep{Karovicova2011}. 
The modeling approach consisted of a radiative transfer model of the 
dust shell using the Monte-Carlo radiative transfer 
code mcsim\_mpi \citep{Ohnaka2006}, which allows us to use
two dust shells with different parameters.
The central stellar source was described by the dust-free 
dynamic model atmosphere P and M series by \citet{Ireland2004b,Ireland2004a}.
The P and M series differ with respect to
the mass of the so-called “parent star”, which is the hypothetical 
non-pulsating equivalent of the pulsating Mira variable.
The model series have been constructed to match the prototype oxygen-rich Mira
stars $o$ Cet and R Leo \citet[for more details see][]{Wittkowski2007}. Following the successful description of IRAS data of a number of Mira
stars by \citet{Lorenz-Martins2000}, we used one dust shell of
Al$_2$O$_3$ grains and one of silicate grains. As in our previous studies,
we used the optical properties of alumina Al$_2$O$_3$ grains 
from \citet{Koike1995} for $\lambda < 8$\,$\mu$m and porous amorphous 
Al$_2$O$_3$ from \citet{Begemann1997} for $\lambda > 8$\,$\mu$m.
For the silicate shell, we used the warm oxygen-deficient astronomical 
silicates by \citet{Ossenkopf1992}, representative of circumstellar 
silicate dust, which include inclusions of metallic iron
and iron oxides. The size of dust grains was set to 0.1 $\mu$m for all grains.

As outlined in the introduction (Sect.~\ref{sec:intro}), the dust contents
and the dust condensation process of oxygen-rich AGB stars is currently
being heavily debated and is not yet understood. It is likely that other 
grain species and other grain sizes than those included in our modeling
approach are present in the circumstellar environments of Mira variables
\citep[e.g.][]{Gail1999}. 
However, since this mix of Al$_2$O$_3$ and silicate grains was successful to 
describe both the IRAS spectra \citep{Lorenz-Martins2000} and the MIDI
flux and visibility spectra \citep{Wittkowski2007,Karovicova2011} of a
number of Mira variables, we continued to use this modeling approach
for the purpose of a characterization of the dust shell geometry and 
optical depth. In order to minimize the (already plentiful) number of fit 
parameters, we refrained from adding additional grain 
properties in the fit procedure. A few 
first attempts were recently made to include dust formation in dynamic model 
atmosphere calculations in a self-consistent way for oxygen-rich AGB stars 
\citep{Ireland2006,Ireland2008,Ireland2011,Sacuto2013}. 
However, as the description of the dust formation is not yet definite, 
we also refrained from a direct comparison of our MIDI data to such 
dynamic model atmospheres with dust formation included.
A more detailed discussion on the dust condensation process follows
below in Sect.~\ref{sec:dust}.

Our fit procedure was the same as described by \citet{Karovicova2011}:
We used ten dynamic atmosphere models from the M series 
that are expected to be better-suited to 
describe our stars. The models cover one complete cycle of 
the M series by \citet{Ireland2004b,Ireland2004a}: M16n (model visual
phase $\Phi_\mathrm{model}=0.60$), M18 (0.75), M18n (0.84), M19n
(0.90), M20 (0.05), M21n (0.10), M22 (0.25), M23n (0.30), M24n (0.40),
and M25n (0.50). For each of these models, we computed a grid of 
radiative transfer models of the dust shell. The radiative transfer
model included variations of 6 parameters, 3 for each of the independent
Al$_2$O$_3$ and silicate shells, the optical depths 
$\tau_\mathrm{V}$ (Al$_2$O$_3$) and $\tau_\mathrm{V}$(silicate), the inner boundary 
radii $R_\mathrm{in}/R_\mathrm{Phot}$ (Al$_2$O$_3$) and
$R_\mathrm{in}/R_\mathrm{Phot}$ (silicate), and the density gradients
$p_\mathrm{A}$ (Al$_2$O$_3$) and $p_\mathrm{B}$ (silicate). 
The first grid included all combinations of optical depths $\tau_\mathrm{V}$
(Al$_2$O$_3$) = 1.0, 1.3, 1.6, 1.9, 2.1, 2.4, 2.7, 3.1; 
$\tau_\mathrm{V}$ (silicate) = 0.5, 1.0, 1.5, 2.0,
2.5, 3.0, 3.5, 4.0, 4.5, 5.0; $R_\mathrm{in}/R_\mathrm{Phot}$
(Al$_2$O$_3$) = 2.0, 2.5, 3.0; $R_\mathrm{in}/R_\mathrm{Phot}$
(silicate) = 2.5, 3.5, 4.5, 5.5, 6.5; $p_\mathrm{A}$ (Al$_2$O$_3$) =
2.0, 2.5, 3.0, 3.5; and $p_\mathrm{B}$ (silicate) = 2.0, 2.5, 3.0,
3.5. In a following step, a grid with finer steps around the
parameters with lowest $\chi^2$ value was
computed. For each individual fit, the angular diameter of the atmosphere
model $\Theta_\mathrm{Phot}$ was the only free fit parameter. 
Here, the photospheric radius $R_\mathrm{Phot}$ is defined as the 1.04\,$\mu$m 
continuum radius of the P/M model atmospheres. 
We remind the reader that we weighted down the part of the photometric 
spectra around 9.5\,$\mu$m, which is strongly affected by telluric absorption. 
The weight of all other data points was given by the corresponding uncertainty.
From all fits, we selected the results with the best $\chi^2$ values. 
It should be mentioned that with the available accuracy, 
a few models of the final selection fit equally well.

\begin{figure*}
\centering
\includegraphics[height=0.35\textheight,angle=90]{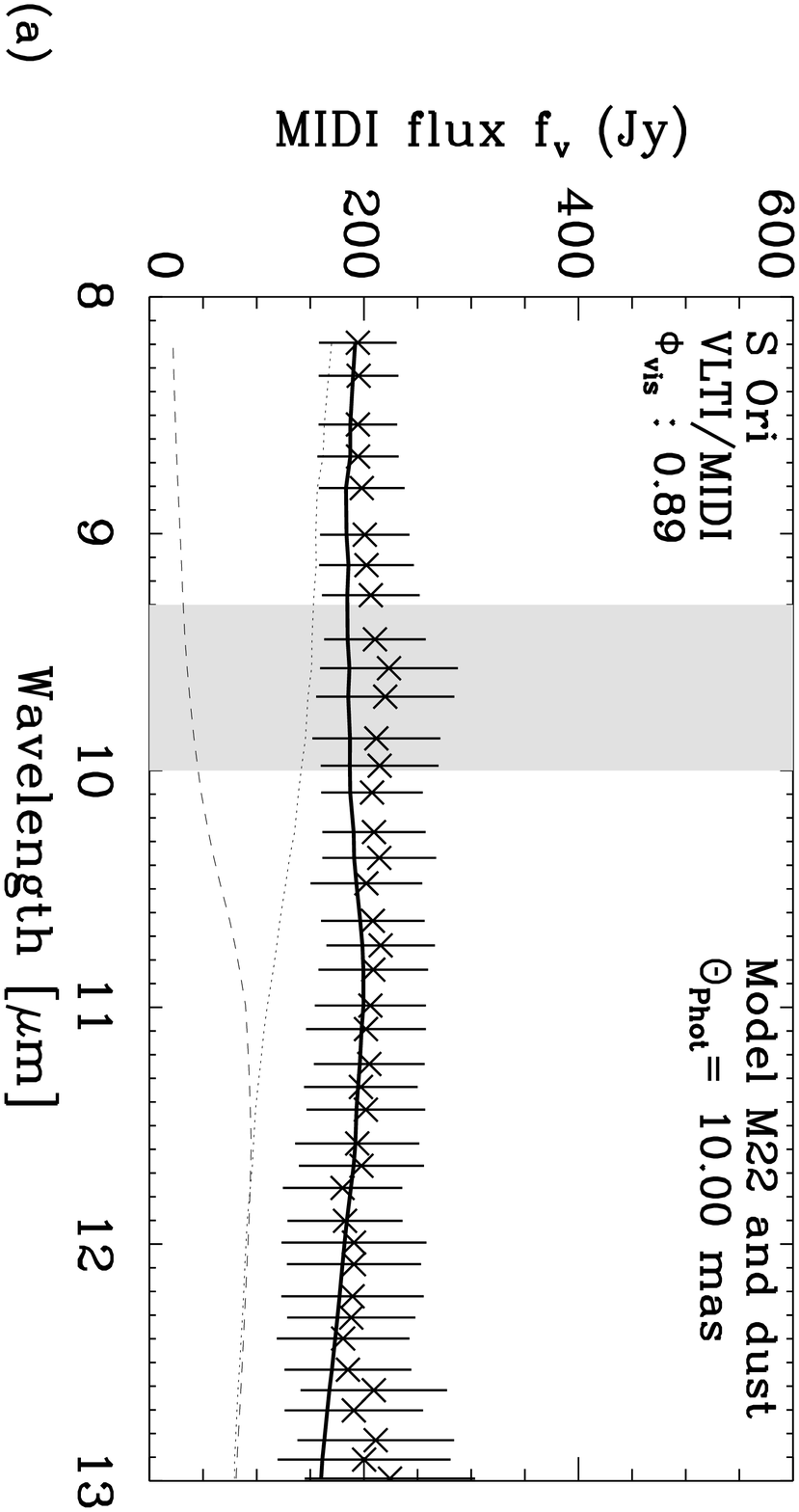}
\includegraphics[height=0.35\textheight,angle=90]{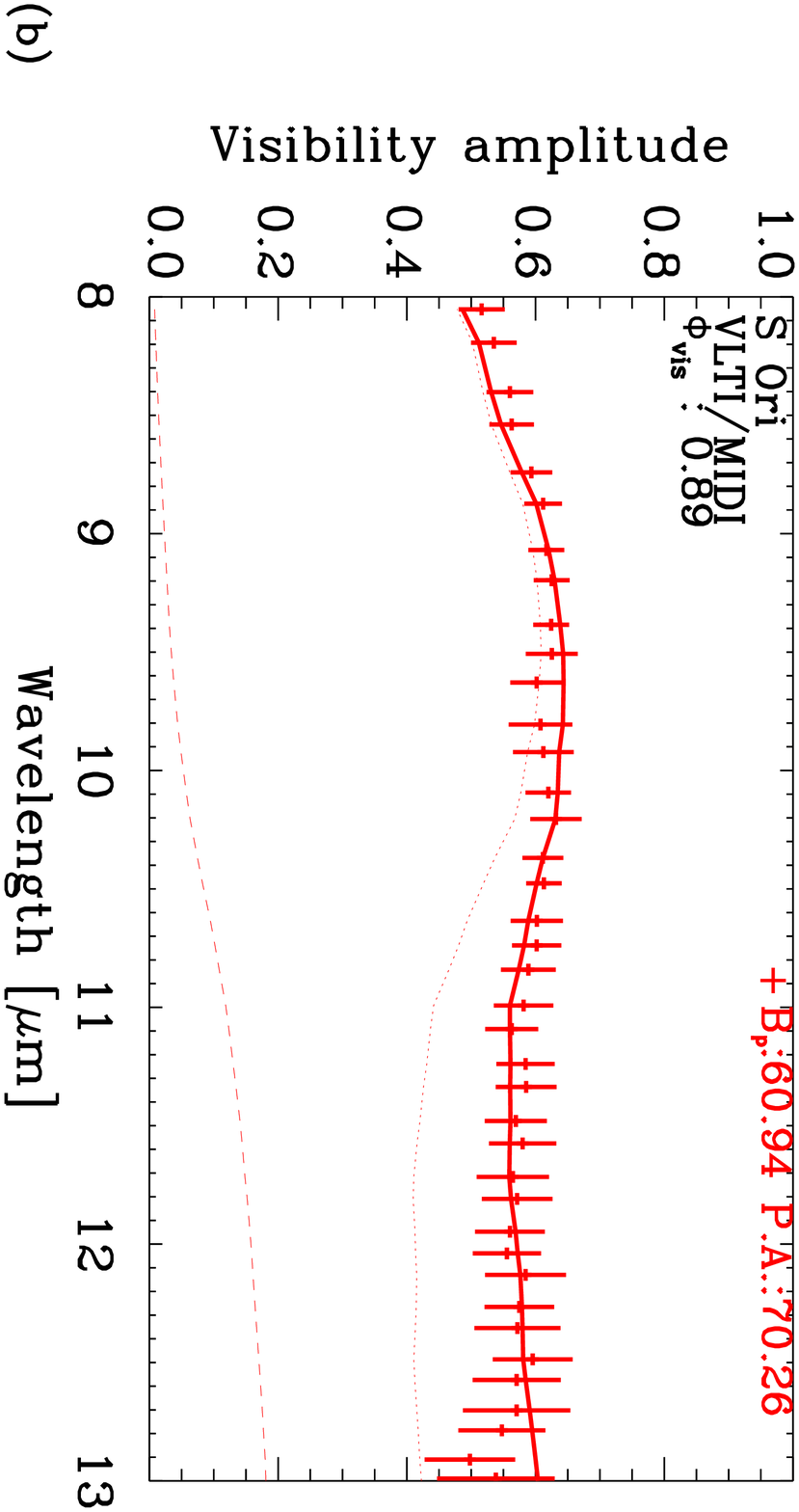}
\includegraphics[height=0.35\textheight,angle=90]{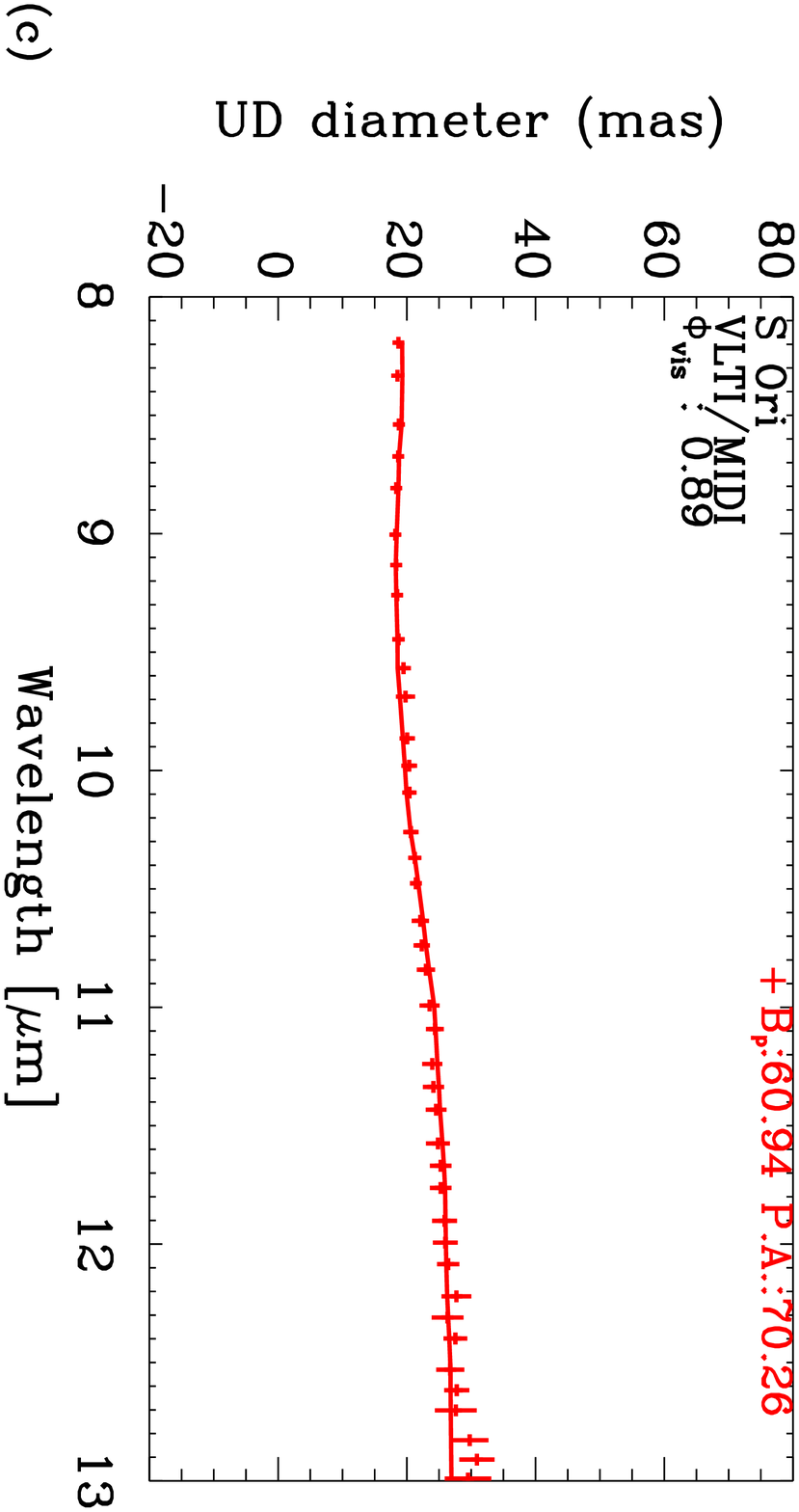}
\includegraphics[height=0.35\textheight,angle=90]{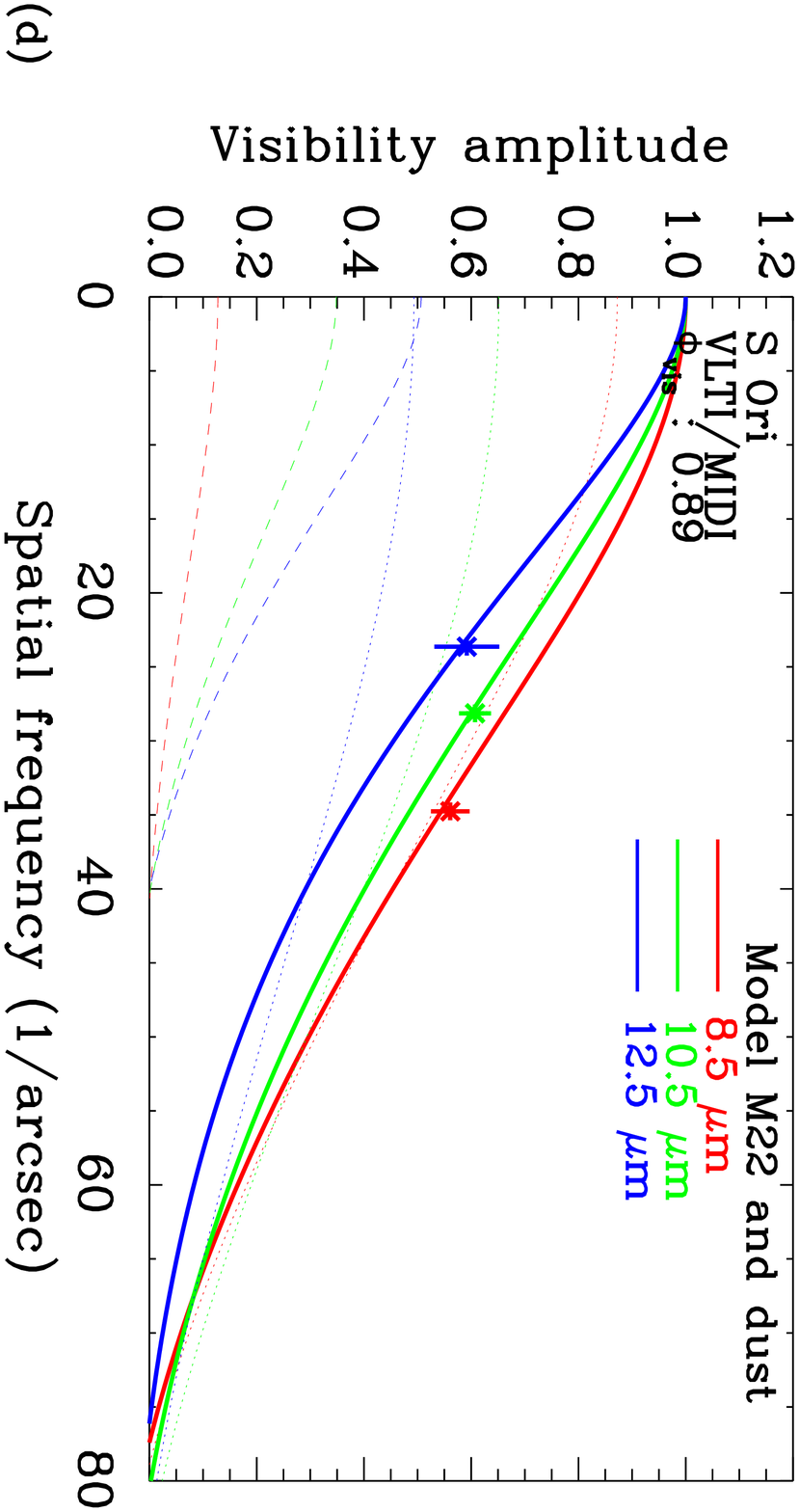}
\caption[S~Ori epoch B]{VLTI/MIDI interferometry at 8-13~$\mu$m of
  S~Ori for the example of epoch B (stellar phase 0.89 see Tab.\ref{observations_sori}).
  The panels show (upper left) the flux, (upper right) the
  visibility amplitude, (lower left) the corresponding UD diameter, 
  and (lower right) the visibility amplitude as a function of spatial frequency
  for three averaged band passes of 8-9 $\mu$m, 10-11 $\mu$m, and 12-13
  $\mu$m. 
  The gray shade indicates the wavelength region around
  9.5\,$\mu$m that is affected by atmospheric absorption. 
  The crosses with error bars denote the measured values. The
  solid lines indicate our best-fit model. 
  The contributions of the stellar and dust components
  alone are indicated by the dotted and the dashed line,
  respectively.}
\label{sori_B}
\end{figure*}

\begin{figure*}
\centering
\includegraphics[height=0.35\textheight,angle=90]{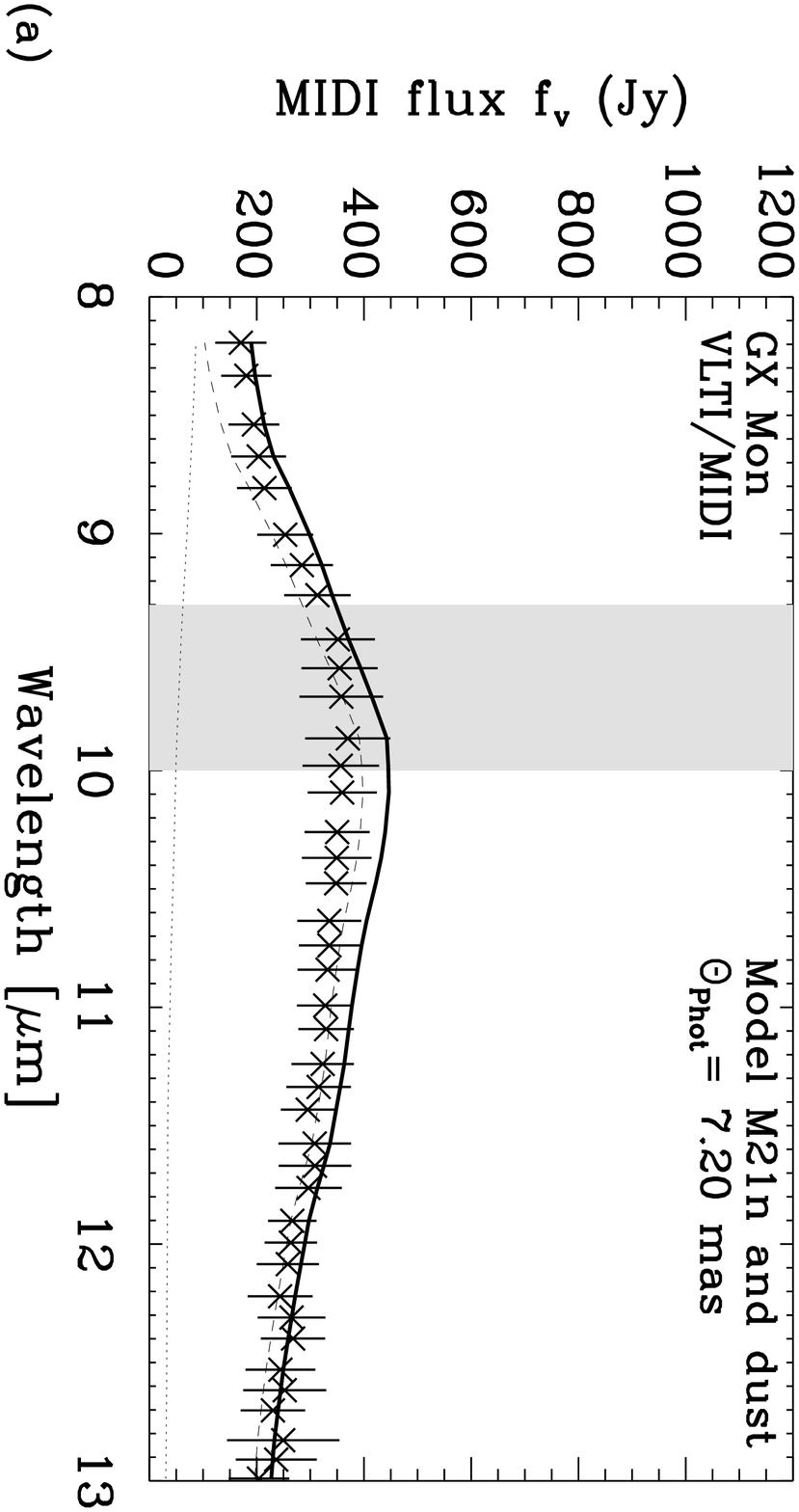}
\includegraphics[height=0.35\textheight,angle=90]{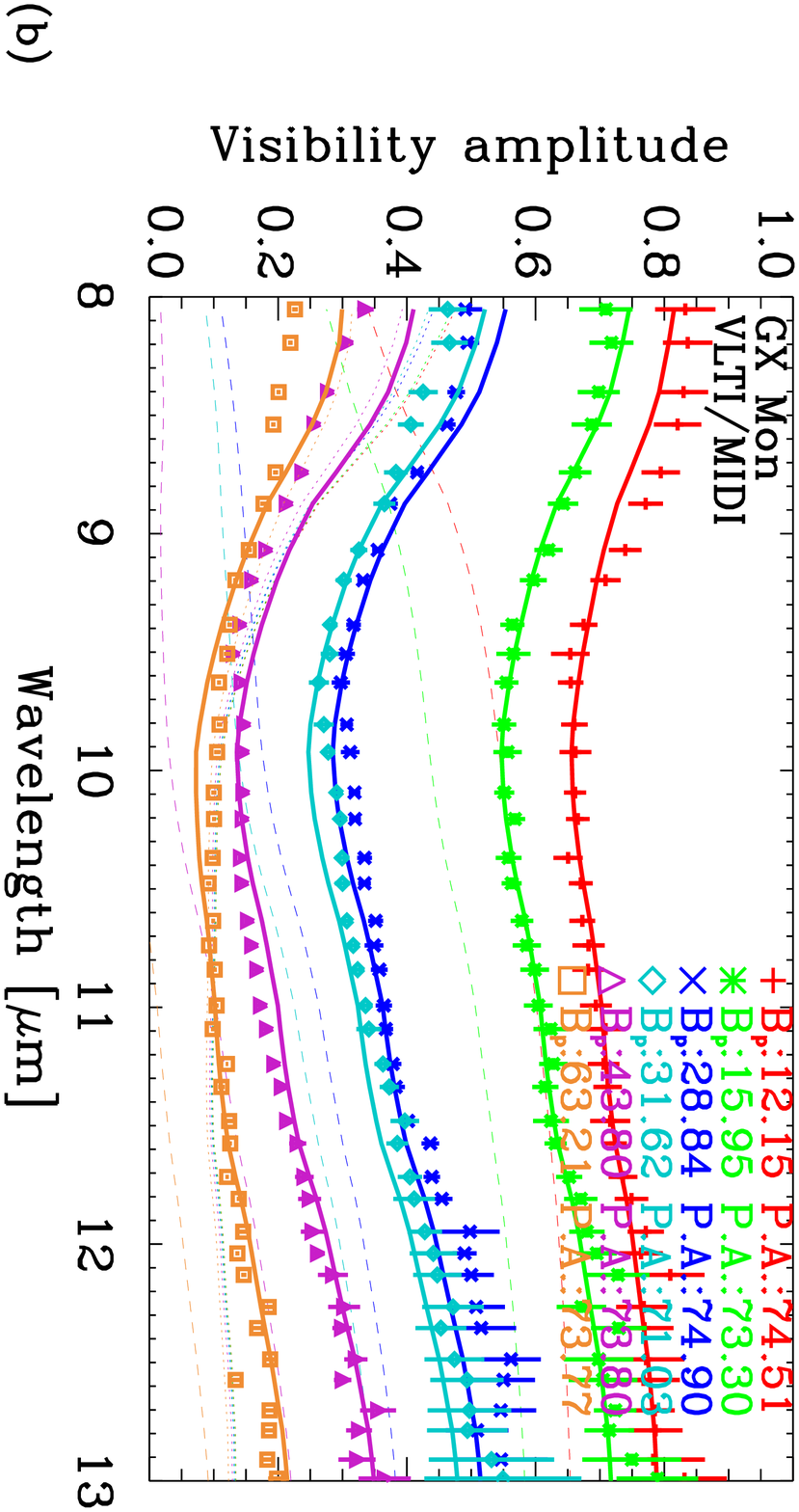}
\includegraphics[height=0.35\textheight,angle=90]{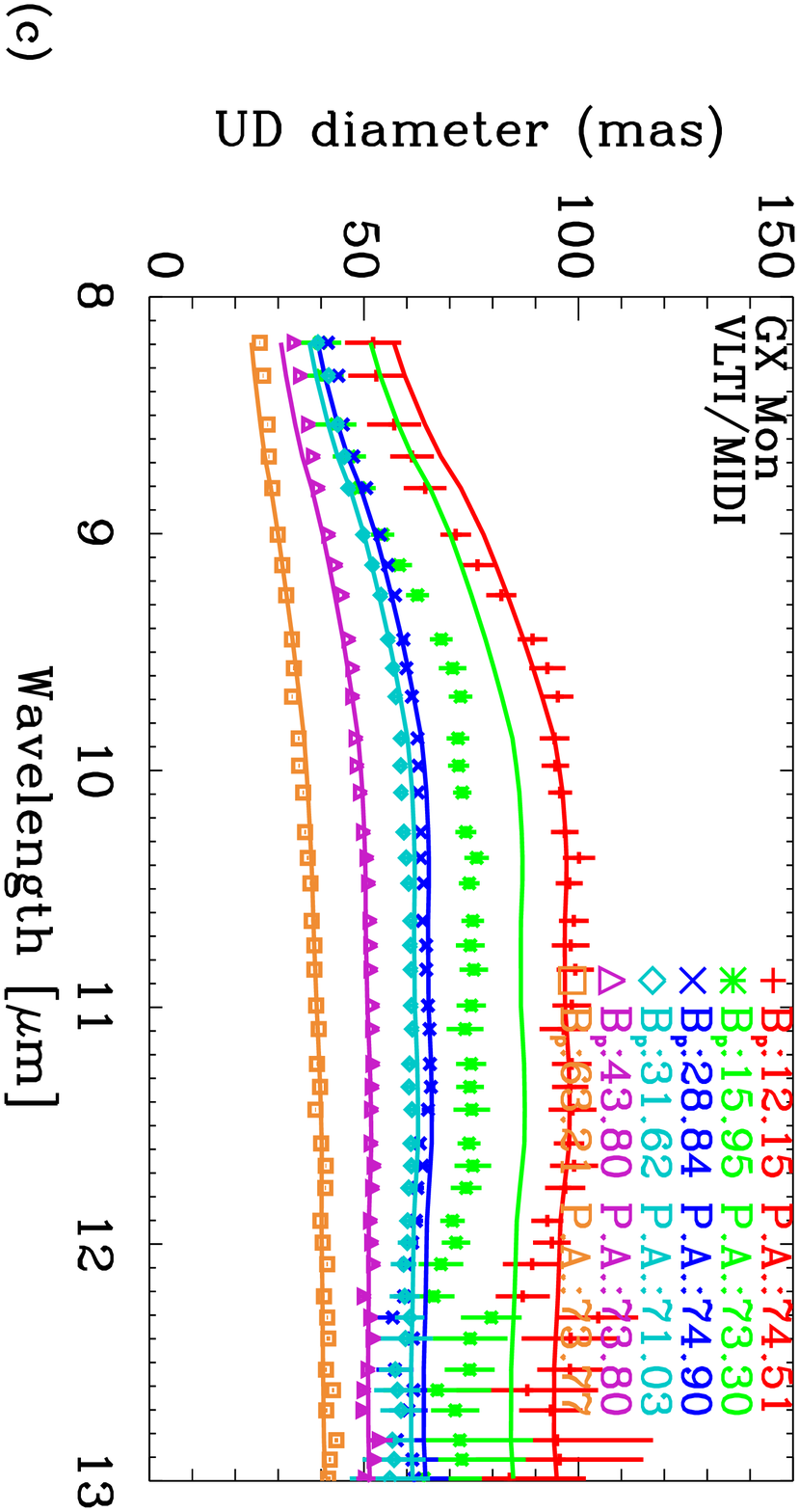}
\includegraphics[height=0.35\textheight,angle=90]{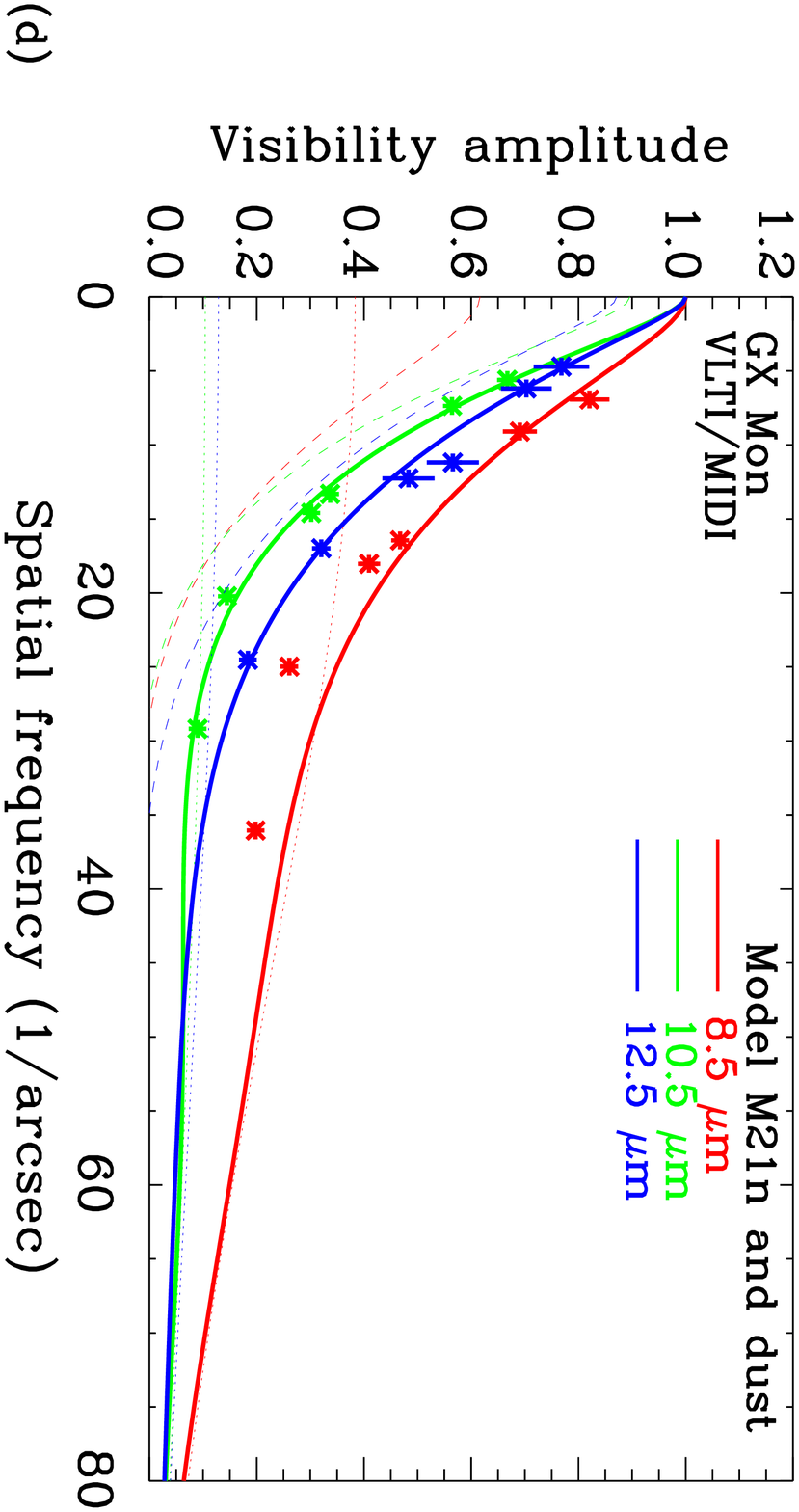}
\caption[GX~Mon epoch E]{As Fig.~\protect\ref{sori_B}, but for the example
of epoch E of GX~Mon (see Tab.~\ref{observations_gxmon}).}
\label{gxmon_E}
\end{figure*}

\begin{figure*}
\centering
\includegraphics[height=0.35\textheight,angle=90]{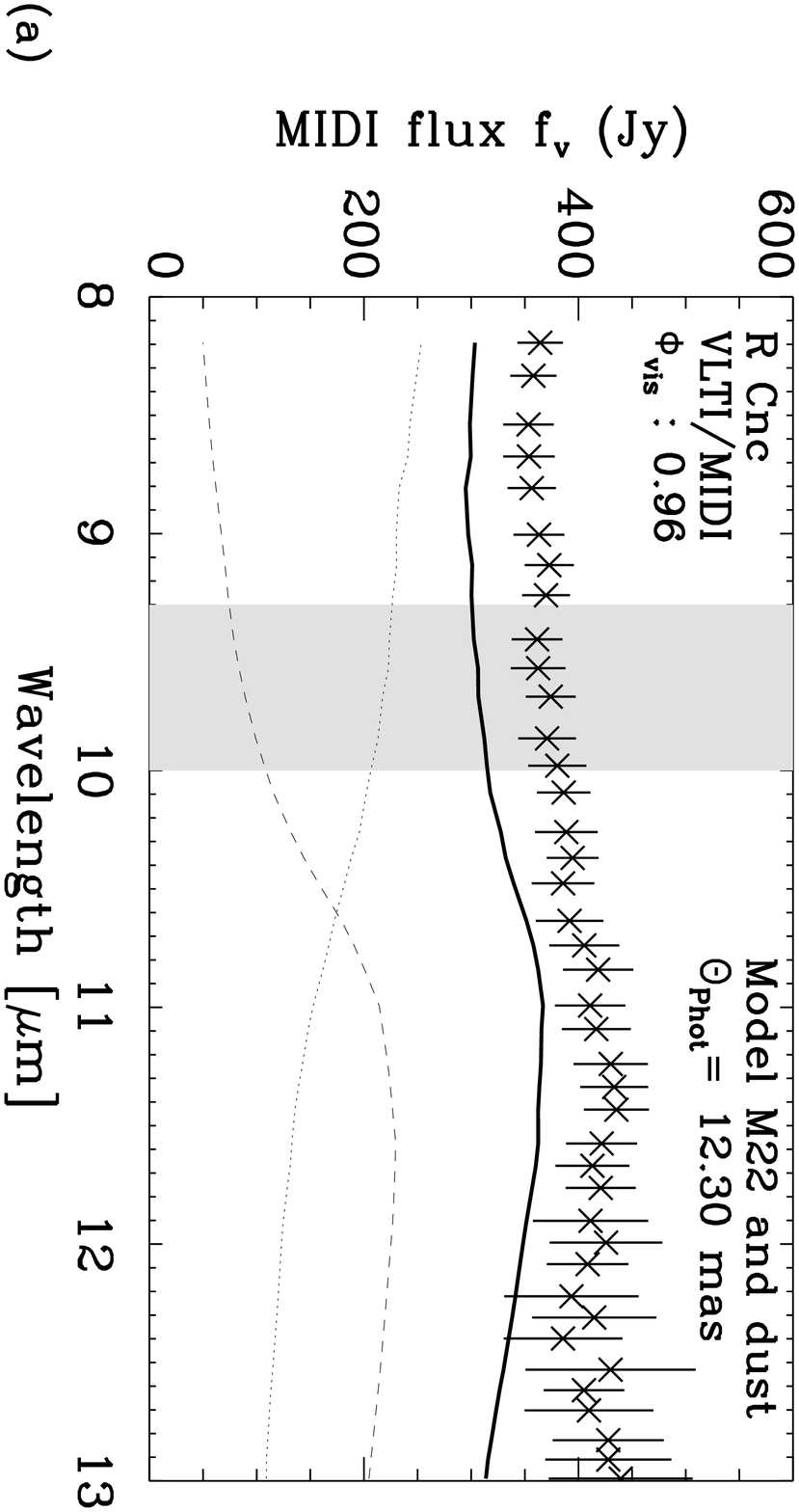}
\includegraphics[height=0.35\textheight,angle=90]{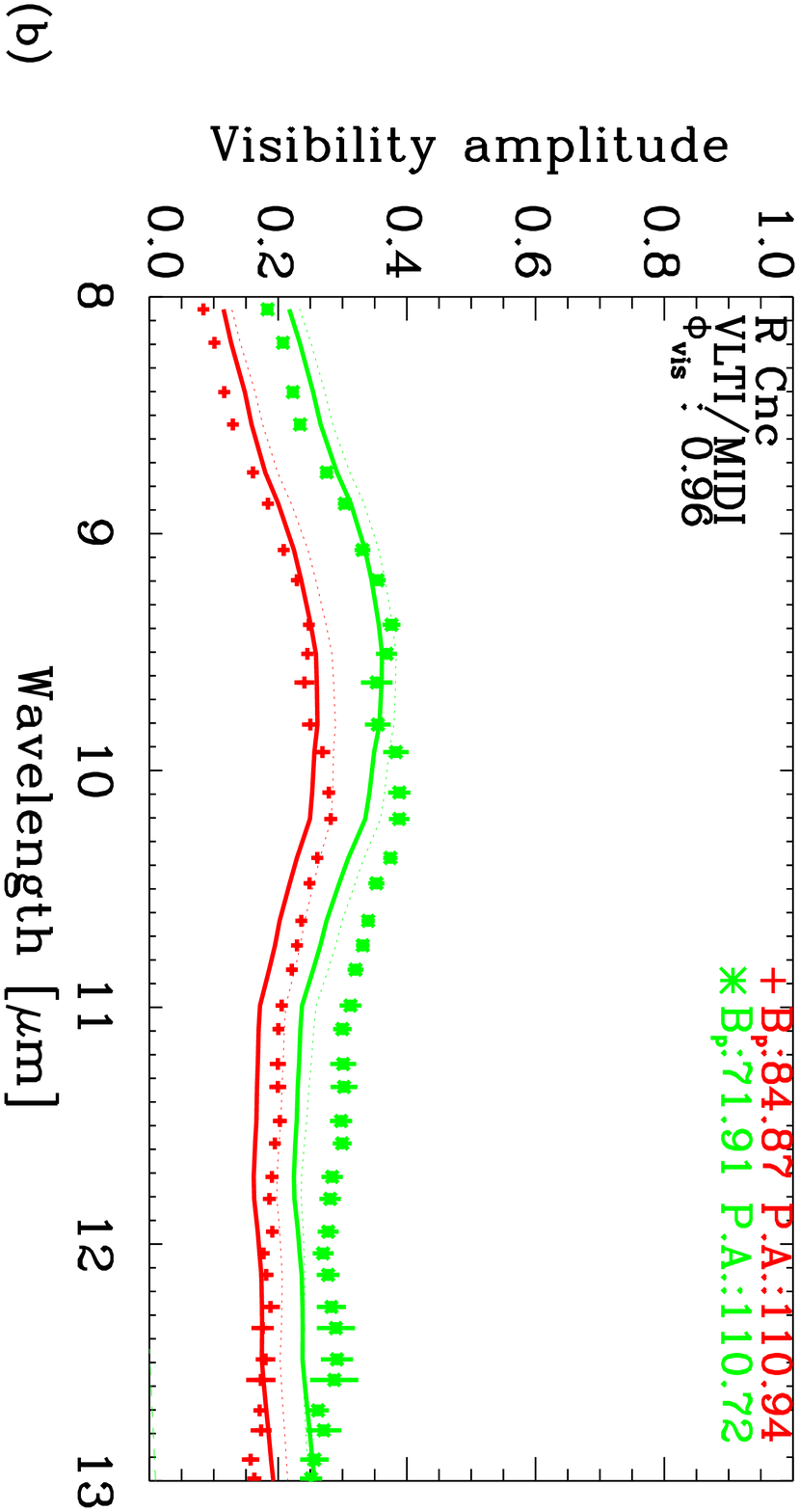}
\includegraphics[height=0.35\textheight,angle=90]{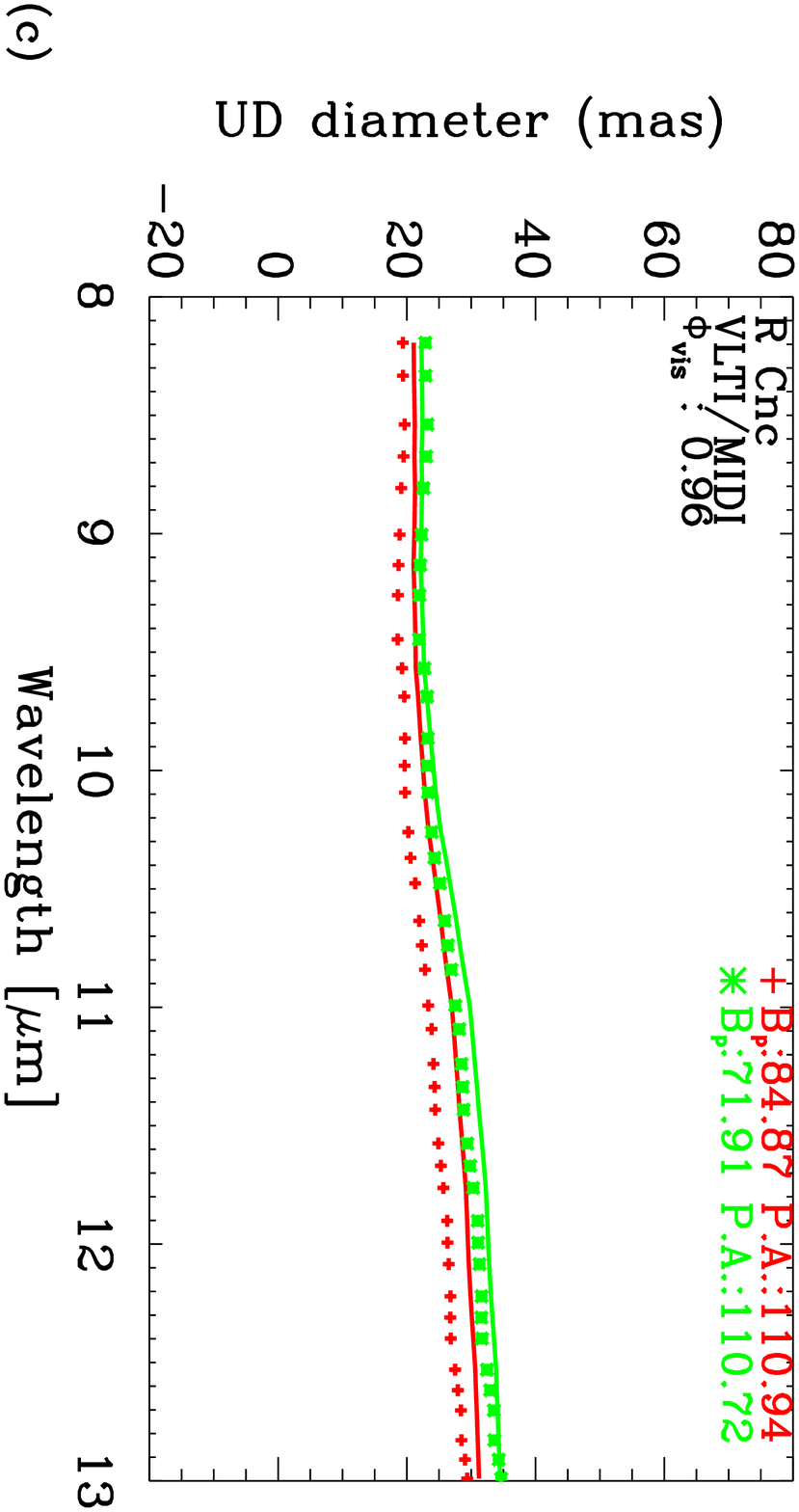}
\includegraphics[height=0.35\textheight,angle=90]{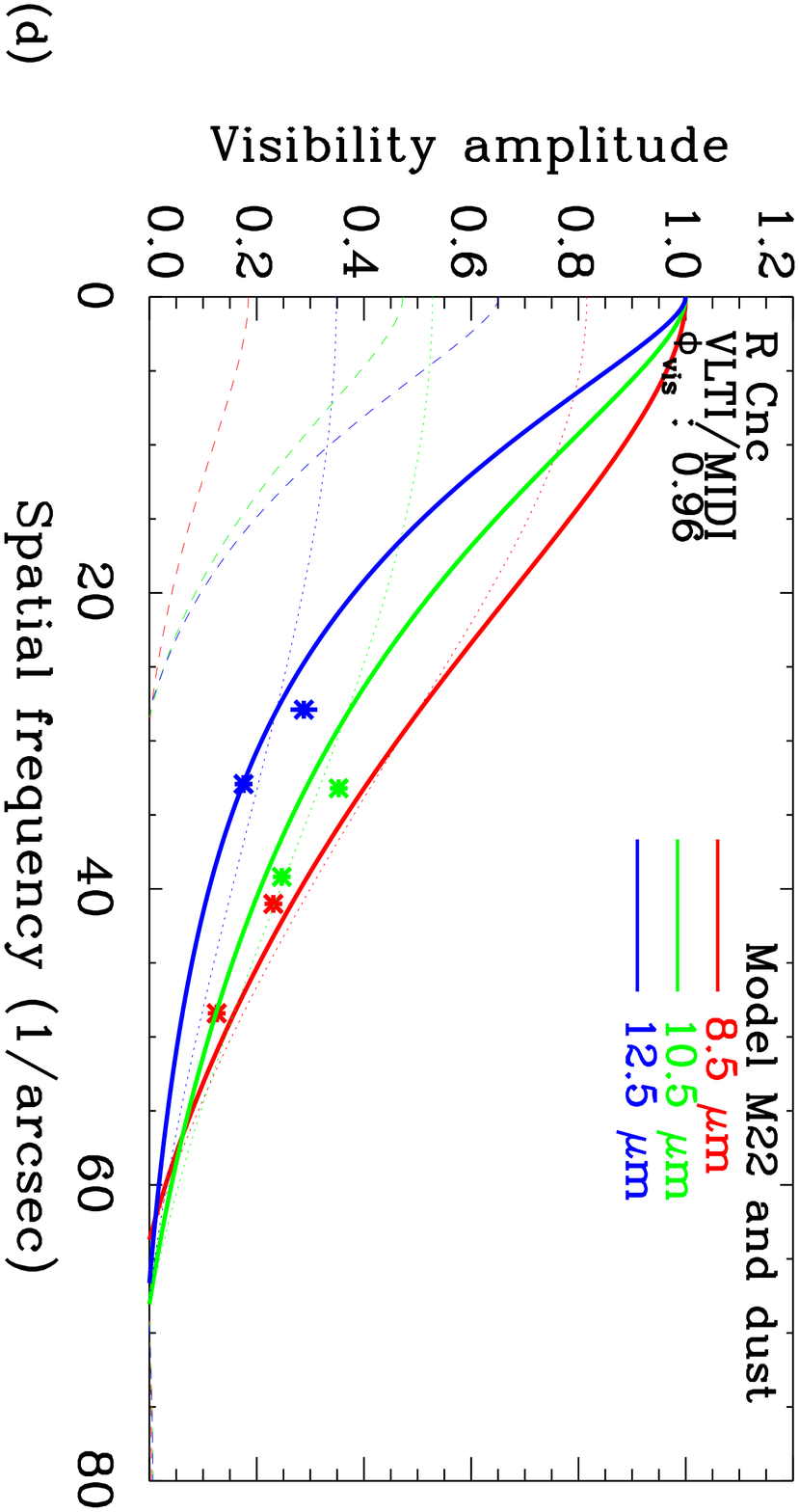}
\caption[R~Cnc epoch A]{As Fig.~\protect\ref{sori_B}, but for the example
of epoch A of R~Cnc (see Tab.~\ref{observations_rcnc}).}
\label{rcnc_A}
\end{figure*}

S~Ori and R~Cnc were a part of the study by \citet{Lorenz-Martins2000} 
and were modeled using Al$_2$O$_3$ alone. Consistently, our MIDI spectra 
of S~Ori and R~Cnc do not exhibit any prominent silicate feature.
GX~Mon was not a part of their study. Our MIDI spectra of GX~Mon
show a pronounced silicate feature, indicating that silicate is one
of the components of the dust shell.
Independent of this information, we considered the full grid of models,
including Al$_2$O$_3$ and silicate shells, for all targets.

\section{Results}
\label{sec:midiresults}

\subsection{Best-fit model parameters}
\label{sec:bestfit}

The best-fit results for S~Ori and R~Cnc were obtained with an 
Al$_2$O$_3$ shell without a contribution by silicate dust, consistent
with \citet{Lorenz-Martins2000}. 
The best-fit results 
for GX~Mon were obtained with a silicate shell together with a
contribution by an Al$_2$O$_3$ shell.
Tables~\ref{tab:par_sori}--~\ref{tab:par_gxmon} list the 
best-fit parameters for each epoch of our S~Ori, R~Cnc, and GX~Mon
observations separately.
The tables lists the epoch, the variability phase, the atmosphere model
and its model phase, the optical depth $\tau_\mathrm{V}$, the inner boundary radius
$R_\mathrm{in}/R_\mathrm{Phot}$, the density distribution $p$, and the
continuum photospheric angular diameter $\Theta_\mathrm{Phot}$.
Since the best-fit models for S~Ori and R~Cnc were obtained without a 
contribution by a silicate shell, only the parameters of the Al$_2$O$_3$ 
dust shell are listed for these targets.

\begin{table*}
\caption{Best-fit model parameters for each epoch of our S~Ori observations}
\centering 
\begin{tabular}{ l r l r r r r r}
\hline\hline 
Epoch & $\Phi_\mathrm{vis}$ &Model& $\Phi_\mathrm{mod}$ & $\tau_\mathrm{V}$  & $R_\mathrm{in}/R_\mathrm{Phot}$ &
$p$     & $\Theta_\mathrm{Phot}$ \\
 &  & &            & (Al$_2$O$_3$) &      (Al$_2$O$_3$)        &    (Al$_2$O$_3$)  &[mas]           \\\hline
A &0.40& M24n &0.40& 2.5 &  2.0 &  2.4 &8.5 \\
B &1.89& M22  &0.25& 1.0 &  1.8 &  3.3 &10.0 \\
C &1.95& M21n &0.10& 1.4 &  2.3 &  3.7 &8.8 \\
D &2.01& M21n &0.10& 1.5 &  2.4 &  3.0 &8.9 \\
E &2.10& M22  &0.25& 1.4 &  1.8 &  3.0 &10.5\\
F &2.16& M23n &0.30& 1.1 &  2.2 &  3.6 &10.1 \\
G &2.22& M23n &0.30& 1.2 &  1.8 &  2.7 &10.1 \\
H &2.29& M25n &0.50& 1.2 &  1.8 &  2.5 &8.5 \\
I &2.91& M22  &0.25& 1.4 &  1.7 &  3.0 &10.2 \\
J &3.00& M22  &0.25& 1.2 &  1.8 &  2.5 &11.1 \\
K &3.13& M25n &0.50& 2.4 &  1.5 &  3.0 &9.0 \\
L &3.19& M23n &0.30& 1.8 &  2.2 &  2.5 &10.9\\
M &3.82& M22  &0.25& 1.4 &  1.7 &  2.9 &10.8\\
N &3.96& M25n &0.50& 1.5 &  2.1 &  2.5 &8.7\\
\hline
\end{tabular}
\tablefoot{The table lists the star, the epoch, the phase at the epoch, the
optical depth $\tau_\mathrm{V}$, the inner boundary radius
$R_\mathrm{in}/R_\mathrm{Phot}$, the density distribution $p$, and the
continuum photospheric angular diameter $\Theta_\mathrm{Phot}$.}
\label{tab:par_sori}
\end{table*}

\begin{table*}
\caption{Best-fit model parameters for each epoch of our R~Cnc observations}
\centering 
\begin{tabular}{ l r l r r r r r}
\hline\hline 
Epoch & $\Phi_\mathrm{vis}$ &Model& $\Phi_\mathrm{mod}$ & $\tau_\mathrm{V}$  & $R_\mathrm{in}/R_\mathrm{Phot}$ &
$p$     & $\Theta_\mathrm{Phot}$ \\
 &  & &            & (Al$_2$O$_3$) &      (Al$_2$O$_3$)        &    (Al$_2$O$_3$)  &[mas]           \\\hline
A &4.95& M22 &0.25& 1.5 &  2.0 &  2.5 &12.3 \\
B &5.11& M22 &0.25& 1.2 &  2.4 &  2.5 &12.3 \\
\hline
\end{tabular}
\label{tab:par_rcnc}
\end{table*}

\begin{table*}
\caption{Best-fit model parameters for each epoch of our GX~Mon observations}
\centering 
\begin{tabular}{ l r r r r r r r r r r}
\hline\hline 
Ep.  &Model& $\Phi_\mathrm{mod}$ & $\tau_\mathrm{V}$ & $\tau_\mathrm{V}$  & $R_\mathrm{in}/R_\mathrm{Phot}$   & $R_\mathrm{in}/R_\mathrm{Phot}$&$p$&
$p$     & $\Theta_\mathrm{Phot}$ \\
 &  &     & (Al$_2$O$_3$)          & (sil.) &      (Al$_2$O$_3$)        &    (sil.)  &     (Al$_2$O$_3$)       &   (sil.)  &   [mas]    \\\hline
\hline
A& M21n& 0.10 &1.5 &3.0   &2.0 &4.8 &3.5 &2.5 &9.1\\
B& M21n& 0.10 &1.3 &3.4   &1.8 &4.8 &3.0 &2.5 &10.5\\
C& M25n& 0.50 &1.4 &3.0   &2.3 &4.8 &3.3 &2.5 &7.3\\
D& M21n& 0.10 &2.5 &3.0   &2.4 &4.5 &3.0 &2.5 &7.2\\
E& M21n& 0.10 &1.8 &3.0   &2.5 &4.6 &2.5 &2.5 &7.2\\
F& M21n& 0.10 &1.8 &3.0   &2.4 &4.3 &3.5 &2.5 &7.5\\
G& M21n& 0.10 &1.2 &3.0   &1.8 &4.8 &2.5 &2.5 &7.5\\
H& M23n& 0.30 &2.8 &4.8   &1.8 &4.8 &2.5 &2.5 &10.0\\
I& M23n& 0.30 &3.1 &3.0   &1.7 &4.8 &2.5 &2.5 &10.6\\
J& M21n& 0.10 &2.2 &3.0   &1.8 &4.4 &2.5 &2.5 &9.5\\
K& M21n& 0.10 &1.7 &3.0   &2.6 &4.4 &2.5 &2.5 &8.8\\
L& M21n& 0.10 &2.1 &2.7   &2.2 &4.6 &2.5 &2.5 &8.8\\
\hline
\end{tabular}
\tablefoot{The phases are uncertain, and therefore they are omitted.}

\label{tab:par_gxmon}
\end{table*}

The agreement between the best-fit models and the observed data is 
satisfactory for all targets and epochs. Figs.~\ref{sori_B}--
\ref{rcnc_A} show for each source one example of the observed and 
synthetic flux and visibility spectra.
The general shape of the S~Ori and R~Cnc spectra are similar and were 
discussed in more detail by \citet{Wittkowski2007}. The spectra of GX~Mon
are similar to those of RR~Aql as discussed by \citet{Karovicova2011}.

\begin{table*}
\caption{Average dust shell parameters}
\centering
\begin{tabular}{l l l l r r r r r r r}
\hline\hline 
Star    & $\overline{\Phi_\mathrm{Vis}}$ & $\overline{\Phi_\mathrm{Mod}}$& Model& $\tau_\mathrm{V}$      & $\tau_\mathrm{V}$   & $R_\mathrm{in}/R_\mathrm{Phot}$ & $R_\mathrm{in}/R_\mathrm{Phot}$ &$p$            &$p$       & $\Theta_\mathrm{Phot}$ \\
        &                &    & & (Al$_2$O$_3$) &(silicate)  & (Al$_2$O$_3$)                   &  (silicate)                     & (Al$_2$O$_3$) &(silicate)&[mas]         \\\hline
S~Ori   & 0.1 $\pm$ 0.2   & 0.3 $\pm$ 0.1&(M22--M24n)      & 1.5 $\pm$ 0.5   & --         & 1.9 $\pm$ 0.3                   & --                             & 2.9 $\pm$ 0.4   & 0.0&  9.7 $\pm$ 1.0     \\
R~Cnc   & 0.0 $\pm$ 0.1   & 0.25&(M22)      & 1.35 $\pm$ 0.2  & --         & 2.2 $\pm$ 0.3                   & --                             & 2.5 $\pm$ 0.0   & 0.0& 12.3 $\pm$ 0.0     \\
 GX~Mon & --             & 0.2 $\pm$ 0.1&(M21n--M23n)    & 1.9 $\pm$ 0.6   & 3.2 $\pm$ 0.5& 2.1 $\pm$ 0.3                   & 4.6 $\pm$ 0.2                   & 2.8 $\pm$ 0.4   & 2.5 $\pm$ 0.0 &8.7 $\pm$ 1.3    \\
RR~Aql$^*$ & 0.6 $\pm$ 0.1 & 0.1&(M21n)    & -- & 2.8 $\pm$ 0.8 & -- & 4.1 $\pm$ 0.7 & -- & 2.6 $\pm$ 0.3 & 7.6 $\pm$ 0.6 \\\hline
\end{tabular}
\tablefoot{The table lists the star, the average observed visual phase $\overline{\Phi_\mathrm{Vis}}$, 
the phase of the model $\overline{\Phi_\mathrm{Mod}}$, the model, the
optical depth $\tau_\mathrm{V}$(Al$_2$O$_3$) and $\tau_\mathrm{V}$(silicate), the inner boundary radius
$R_\mathrm{in}/R_\mathrm{Phot}$(Al$_2$O$_3$) and $R_\mathrm{in}/R_\mathrm{Phot}$(silicate), 
the density distribution $p$(Al$_2$O$_3$) and p(silicate), and the
continuum photospheric angular diameter $\Theta_\mathrm{Phot}$.\\
$^*$: The values for RR~Aql are from \citet{Karovicova2011} and are repeated 
here for completeness.}
\label{tab:par_avg_gxmon}
\end{table*}

\begin{figure}
\centering
\includegraphics[width=0.5\textwidth]{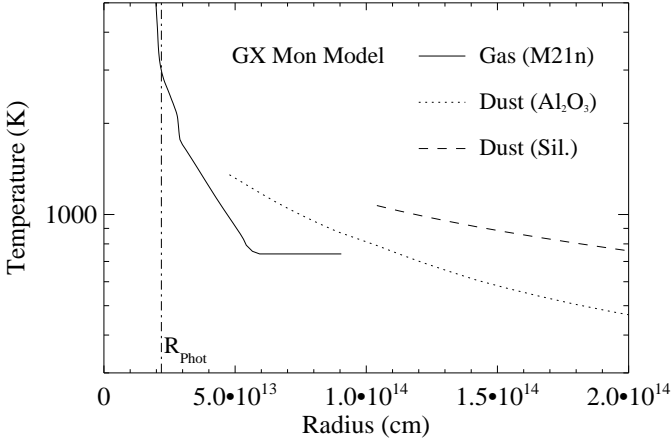}
\caption{Stratification of the gas temperature and dust temperature 
(Al$_2$O$_3$ and silicate dust) for the example of our GX~Mon model.
The solid line shows the gas temperature (model M21n), the dotted
line the dust temperature of the Al$_2$O$_3$ dust shell, and the
dashed line the dust temperature of the silicate dust shell. 
The vertical line indicates $R_\mathrm{Phot}$=2,19\,$\times$\,10$^{13}$\,cm.
For the model parameters, see Tab.~\protect\ref{tab:par_avg_gxmon}.}
\label{fig:temp}
\end{figure}

The model fit results obtained at the individual epochs did not indicate 
any significant dependence of the dust formation on phase or cycle within 
the phase coverage of our observations. 
In addition, a direct comparison of the flux and visibility spectra
(cf. Sect.~\ref{sec:variability} below) obtained at different phases
and cycles showed a good agreement and no variability within our
measurement uncertainties.
As a result, we concentrate in 
the following on the average best-fit dust shell parameters. 
Tab. \ref{tab:par_avg_gxmon} lists these averaged dust shell
parameters together with the averaged observational phases.
Note that we do not cover a full cycle, therefore we indicate 
the average observational phase and the most frequent model.
The averaged model parameters belong to a 
certain averaged phase, and may be different for a 
different phase.
Fig.~\ref{fig:temp} shows the stratification of the gas
and dust temperatures for the example of the average GX~Mon 
model, which consists of an Al$_2$O$_3$ and a silicate dust shell.
The dust temperatures at the inner radii are given in the 
paragraphs on the individual sources below.
%MWI As discussed, we could, but we don't have to, add a plot of the
%intensity profile. To be created by MWI or IKA in parallel to the
%review by the co-authors.
\paragraph{S~Ori:}
Our S~Ori observations had an average visual phase
of $\overline{\Phi_\mathrm{Vis}}=0.1\pm0.2$ (post-maximum phase).
The average model phase was $\overline{\Phi_\mathrm{Mod}}=0.3\pm0.1$,
corresponding to model M22/M23n.
The average optical depth of the Al$_2$O$_3$ dust shell was 
$\tau_V$(Al$_2$O$_3$)=1.5$\pm$0.5 at $\lambda$\,=\,0.55\,$\mu$m
(corresponding to 0.04 at $\lambda$\,=\,8\,$\mu$m, 0.29 at
$\lambda$\,=\,12\,$\mu$m, and a maximum within 8--12\,$\mu$m of 0.30
at $\lambda=11.8$\,$\mu$m). The inner radius of the Al$_2$O$_3$ dust 
shell was 
$R_\mathrm{in}$\,=\,1.9$\pm$0.3\,$R_\mathrm{Phot}$, and the power-law
index of the density distribution was $p$\,=\,2.9$\pm$0.4. The dust
temperature at $R_\mathrm{in}$ was 1340\,K, close to the
condensation temperature of Al$_2$O$_3$ of $\sim$1400\,K \citep{Gail2010}.
The average
photospheric angular diameter resulted in
$\Theta_\mathrm{Phot}$\,=\,9.7$\pm$1.0\,mas. This value is consistent 
with the K-band ($\lambda$\,=\,2.2\,$\mu$m, $\Delta\lambda$\,=\,0.4\,$\mu$m) 
UD diameter values measured to be between 9.6\,mas and 10.5\,mas by
\citet{vanBelle1996}, \citet{Millan-Gabet2005}, and
\citet{Boboltz2005}. 
The best-fit parameters for S~Ori are also consistent with those
derived from the earlier VLTI/MIDI observations by \citet{Wittkowski2007},
who derived at near-minimum and post-maximum phases Al$_2$O$_3$ dust 
shells with inner boundary radii between 1.8\,$R_\mathrm{Phot}$ and 
2.4\,$R_\mathrm{Phot}$ and optical depths between $\tau_\mathrm{V}$(Al$_2$O$_3$)=1.5
(post-maximum) and $\tau_\mathrm{V}$(Al$_2$O$_3$)=2.5 (near minimum).
Our average photospheric angular diameter 
at an average post-maximum visual phase is also consistent
with the post-maximum diameter derived from the previous VLTI/MIDI S~Ori 
observations \citep{Wittkowski2007} of 9.7\,mas at visual phase 0.16.
Our value for $\Theta_\mathrm{Phot}$ corresponds
to a photospheric radius of $R_\mathrm{Phot}$=501$^{+191}_{-164}$R$_{\odot}$.
With the bolometric magnitude of S~Ori of m$_\mathrm{bol}$=3.08
and $\Delta$m$_\mathrm{bol}$=0.49 from \citet{Whitelock2000},
we derived an effective temperature of
T$_\mathrm{eff}\sim$2627$\pm$300K. 

\paragraph{R~Cnc:}
Our R~Cnc observations had an average visual phase of 
$\overline{\Phi_\mathrm{Vis}}=0.0\pm0.1$ (near-maximum phase).
The average model phase was $\overline{\Phi_\mathrm{Mod}}=0.25$,
corresponding to model M22.
The average optical depth of the Al$_2$O$_3$ dust shell was 
$\tau_\mathrm{V}$(Al$_2$O$_3$)=1.35$\pm$0.2 at $\lambda=0.55$\,$\mu$m 
(corresponding to 0.03 at $\lambda=8$\,$\mu$m, 0.27 at $\lambda=12$\,$\mu$m, 
and a maximum within 8--12\,$\mu$m of 0.27 at $\lambda=11.8$\,$\mu$m). 
The inner radius of the dust shell was 
$R_\mathrm{in}$=2.2$\pm$0.3\,$R_\mathrm{Phot}$, 
and the power-law index of the density distribution was $p$=2.5.
The dust temperature at $R_\mathrm{in}$ was 1210\,K,
below the condensation temperature of Al$_2$O$_3$ of $\sim$1400\,K.
The average photospheric angular diameter resulted in 
$\Theta_\mathrm{Phot}$=12.3\,mas,
which is consistent with the value of 11.8\,mas $\pm$ 0.7 mas derived 
from near-infrared VLTI/AMBER observations by \citet{Wittkowski2011} 
at phase 0.3.
Our value for $\Theta_\mathrm{Phot}$ corresponds
to a photospheric radius of $R_\mathrm{Phot}$=371$\pm37$R$_{\odot}$.
Together with the bolometric magnitude of m$_\mathrm{bol}$=2.60
and $\Delta$m$_\mathrm{bol}$=1.05 from \citet{Whitelock2000},
we derived an effective temperature of
T$_\mathrm{eff}\sim$2604$\pm$300K. 

\paragraph{GX~Mon:}
The visual phase of our GX~Mon observations is unknown.
The average model phase was $\overline{\Phi_\mathrm{Mod}}=0.2\pm0.1$,
corresponding to model M21n/M22.
An Al$_2$O$_3$ dust shell in combination with a
silicate dust shell provided the best agreement with our data. The
average optical depth of the Al$_2$O$_3$ dust shell was 
$\tau_\mathrm{V}$(Al$_2$O$_3$)=1.9$\pm$0.6 and the average optical depth of the
silicate dust was $\tau_\mathrm{V}$(silicate)=3.2$\pm$0.5 at
$\lambda=0.55$\,$\mu$m corresponding to 0.05 at $\lambda$=8\,$\mu$m (Al$_2$O$_3$) 
and 0.03  at $\lambda$=8\,$\mu$m (silicate), 0.38 at $\lambda$=12\,$\mu$m (Al$_2$O$_3$) 
and 0.09 at $\lambda$=12\,$\mu$m (silicate). The inner radius of the Al$_2$O$_3$ dust shell
was $R_\mathrm{in}$(Al$_2$O$_3$)=2.1$\pm$0.3\,$R_\mathrm{Phot}$ and the
inner radius of the silicate dust shell was 
$R_\mathrm{in}$(silicate)=4.6$\pm$0.2\,$R_\mathrm{Phot}$. The
power-law indices of the density distribution were
$p$(Al$_2$O$_3$)=2.8$\pm$0.4 and $p$(silicate)=2.5$\pm$0.0. 
The dust temperatures at the inner radii of the Al$_2$O$_3$
and silicate shells were 1350\,K and 1070\,K, respectively, 
consistent with the typical condensation temperatures of 
$\sim$1400\,K and $\sim$1000\,K \citep{Gail2010}.
The average photospheric angular diameter resulted in
$\Theta_\mathrm{Phot}$=8.7$\pm$1.3\,mas.
A previous direct diameter measurement for GX~Mon is not available.
However, our best-fit angular diameter is consistent with
the estimate based on the surface brightness calibration 
by \citet{vanBelle1999} of 6--7.5\,mas.
For our value for $\Theta_\mathrm{Phot}$ we derived
a photospheric radius of $R_\mathrm{Phot}$=675$^{+300}_{-249}$R$_{\odot}$.
A bolometric magnitude of m$_\mathrm{bol}$=4.09
and M$_\mathrm{bol}$=-5.19 was adopted from \citet{Olivier2001}.
These values correspond to an effective temperature of
T$_\mathrm{eff}\sim$2173$\pm$300K.

\paragraph{RR~Aql:}
The dust shell parameters for RR~Aql were already reported 
by \citet{Karovicova2011}.
We repeat the average best-fit parameters, because we will discuss the
general dust shell properties based on all 4 targets in the following.
RR~Aql could best be described with a silicate dust shell alone, consistent
with \citet{Lorenz-Martins2000}.
The average visual phase was $\overline{\Phi_\mathrm{Vis}}=0.6\pm0.1$ 
(near-minimum phase).
%, described best by model phase 0.1 (M21n). 
The average optical depth of the silicate dust shell was
$\tau_\mathrm{V}$(silicate)=2.8$\pm$0.8 at $\lambda=0.55$\,$\mu$m 
(corresponding to 0.03 at $\lambda$=8\,$\mu$m, 0.06 at $\lambda$=12\,$\mu$m, and a maximum of 0.22 at $\lambda$=9.8\,$\mu$m),
The inner radius was $R_\mathrm{in}$=4.1$\pm$0.7\,$R_\mathrm{Phot}$,
and the power-law index of the density distribution was $p$=2.6.
The dust temperature at the inner radius was 1130\,K, consistent with
typical silicate condensation temperatures.  
The angular diameter corresponds to a photospheric radius
of $R_\mathrm{Phot}$=522$^{+230}_{-140}$R$_{\odot}$
and an effective temperature of T$_\mathrm{eff}$=2420$\pm$200\,K.

\paragraph{Summary:}
The results show that the shape of both the
visibility and the photometry spectra for the Mira variables S~Ori, 
GX~Mon, R~Cnc and RR~Aql, can be well reproduced by the
radiative transfer model of the circumstellar dust shell 
using Al$_2$O$_3$ and silicate dust grains with different inner radii,
where the central source was described by dynamic model atmospheres.
The same approach has already been successful to describe 
IRAS spectra of a number of Mira variables \citep{Lorenz-Martins2000}.
S~Ori and R~Cnc could best be described by only an Al$_2$O$_3$ dust 
shell, RR~Aql by only a silicate dust shell, and GX~Mon by
a combination of an Al$_2$O$_3$ and a silicate dust shell.
The Al$_2$O$_3$ shells have inner radii between 1.9 and 2.2 stellar
photospheric radii, and the silicate shells have inner radii
between 4.1 and 4.6 stellar photospheric radii.
The best-fit photospheric angular diameters are consistent with
independent estimates. The model dust temperatures at the inner
radii of 1.9--2.2 stellar radii and 4.1--4.6 stellar radii are consistent 
with dust condensation temperatures of Al$_2$O$_3$ and silicates,
respectively.

\subsection{Aluminum abundance}
\label{sec:alabundance}
We used an ad-hoc radiative transfer model of the dust shell that
did not self-consistently include the dust formation process
based on the available elements in the extended atmosphere.
However, we investigated whether a sufficient
amount of aluminum is available in the extended atmospheres to match the
required number density of Al$_2$O$_3$ dust grains of our best-fit 
radiative transfer models.
We calculated the aluminum abundances in the gas and the dust based on
our models. We used as an example the best-fit model for S~Ori
consisting of the M22 model atmosphere and an Al$_2$O$_3$ dust shell with
parameters as listed in Tab.~\ref{tab:par_avg_gxmon}. M22 shows a 
relatively compact atmospheric extension, so that we used in addition
model P20 as an example of a more extended atmosphere.

\paragraph{Aluminum number density in the dust component:}
The number density of the dust grains as a function of radius is an output 
of our radiative transfer model, based on the chosen optical depth, the inner
radius, and the power-law index of the density distribution.
We converted the Al$_2$O$_3$ grain number density ($N$)
to the mass density (gcm$^{-3}$) by $4/3\,\times\,\pi\,\times\,a^3\,\times\,\varrho_\mathrm{bulk}\,\times\,N$,
where $a$ is the grain radius, and $\varrho_\mathrm{bulk}$ is the bulk
density of the grain. \citet{Arhammer2011} give bulk densities
of amorphous Al$_2$O$_3$
between 2.1 gcm$^{-3}$ and 3.6 gcm$^{-3}$. We used a mean
value of 2.85 gcm$^{-3}$.
The mass of a single Al$_2$O$_3$ molecule is $102$amu, 
where amu is the atomic mass unit. From the mass density
and this mass of a single Al$_2$O$_3$ molecule we computed the
number density of Al$_2$O$_3$ molecules and thus of
aluminum atoms in the dust shell.

\begin{figure}
  \includegraphics[width=0.48\textwidth]{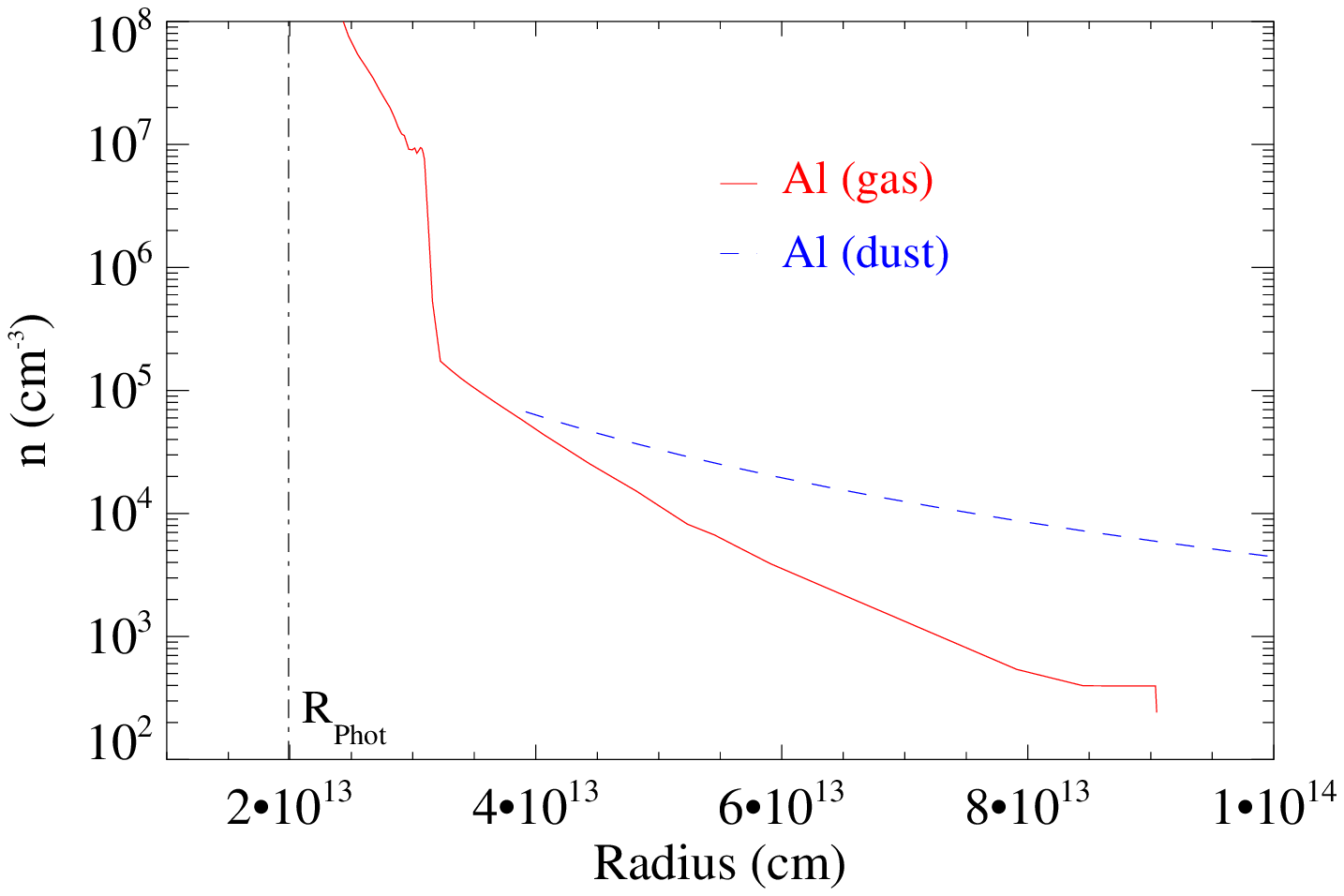}
  \includegraphics[width=0.48\textwidth]{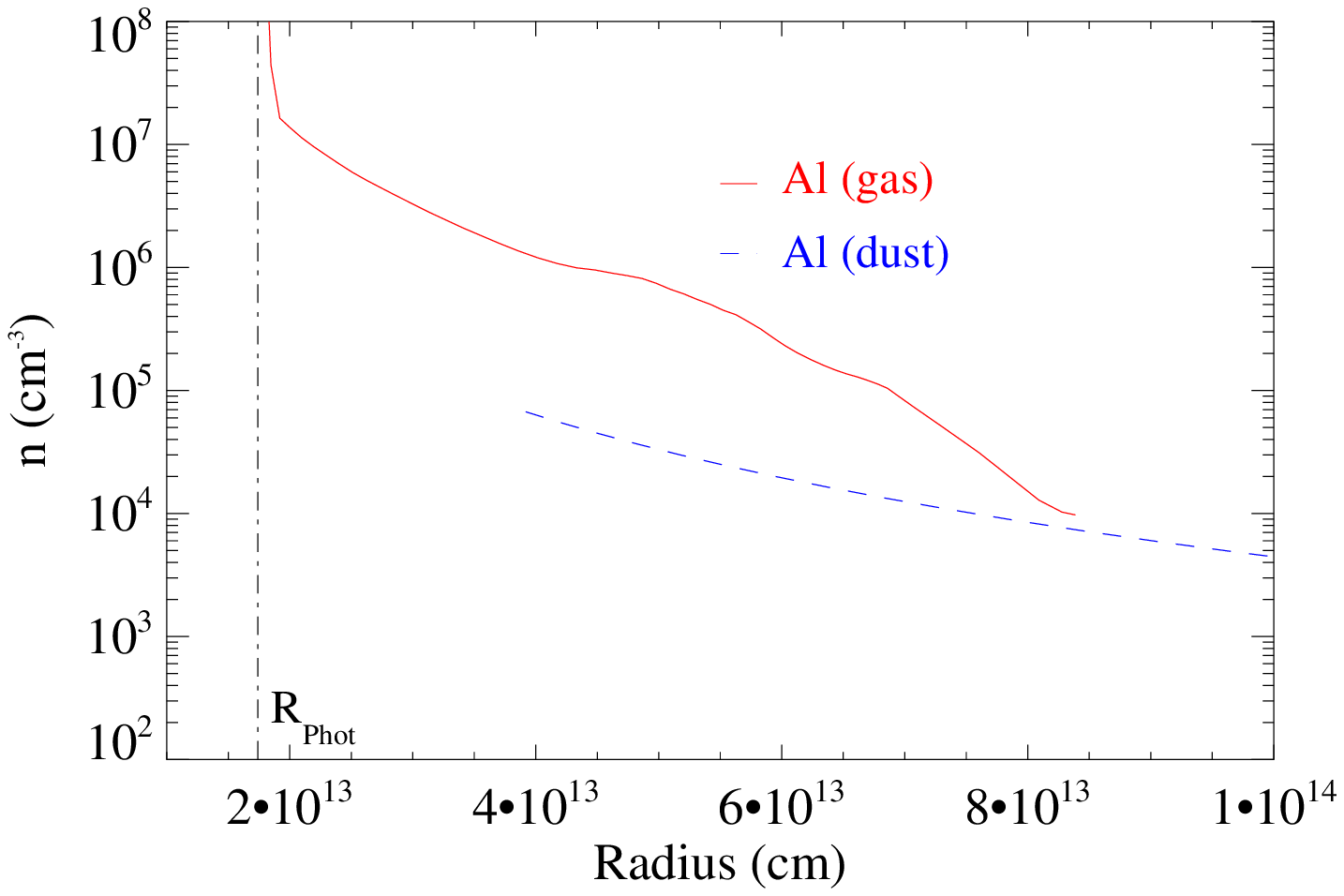}
  \caption{Number density of available aluminum atoms in the dust-free
extended model atmospheres M22 (top) and P20 (bottom) compared
to the number density of aluminum in the dust shell. Shown is the
overlapping radial range between the inner radius of the dust shell
model and the maximum radius of the atmosphere model.
The vertical line indicates $R_\mathrm{Phot}$=1.99\,$\times$\,10$^{13}$\,cm 
(M22) and $R_\mathrm{Phot}$=1.74\,$\times$\,10$^{13}$\,cm (P20).
}
\label{fig:Alcomp}
\end{figure}

\paragraph{Aluminum number density in the extended atmosphere (gas component):}
We calculated the molecular and atomic abundances in
chemical equilibrium for the given temperature and gas stratification
of the (dust free) atmosphere model based on solar abundances.
The total number density of available aluminum atoms was computed
as the sum of all molecules, atoms, and ions containing aluminum.

\paragraph{Comparison of the aluminum number density in the dust and the gas:}
Fig.~\ref{fig:Alcomp} shows the comparison of the aluminum atoms in the 
dust shell to the number of available aluminum atoms in the gas.
Compared to model M22, the number density of aluminum in our dust shell
matches that of the atmosphere at
the inner dust shell radius $R_\mathrm{in}\sim 4\times10^{13}$\,cm.
At larger radii more aluminum atoms are needed for the best-fit
dust shell than available in the atmosphere. This discrepancy
gets wider for larger radii.
Compared to the more extended model P20, a sufficient number
of aluminum atoms is available up to about two times the inner 
dust shell radius, or about 4 times the photospheric radius.
It has to be noted that our best-fit power-law density profile for
the dust shell corresponds to material that is outflowing, while the
gas density is based on a dynamical model without mass loss. This results
in a steeper density for the gas, while the dust density falls off more
gradually. Assuming that the dust grains are accelerated and drag along 
the gas, the gas density would have a smaller gradient as well, and the number
density of aluminum atoms in the gas may match the required numbers
of the dust shell over a larger radial range.

\paragraph{Overall abundance of aluminum:}
Overall, the mass-loss rate at the outer boundary radius of the
dust shell must equal the mass-loss rate just inside the inner boundary.
Just inside $R_\mathrm{in}$, where no dust particles have yet formed,
the mass-loss fraction of aluminum of the gas component is based on the
chemical composition of the gas. Assuming solar metallicity, and using
the number densities by \citet{Asplund2009}, the mass fraction of 
aluminum is $f_\mathrm{Al}(<R_\mathrm{in})=5.5\,\times\,10^{-5}$.
The chemical composition may be altered by hot-bottom-burning and
thermal pulses, but the abundance of aluminum is not expected to change 
significantly and to essentially equal solar system 
abundances \citep{Gail2010}.
In our best-fit models of R~Cnc and S~Ori, the dust shells
consisted only of Al$_2$O$_3$. We thus assumed that
at the outer boundary radius of the dust shell ($R_\mathrm{out}$)
all aluminum is bound in Al$_2$O$_3$, and that this is the only
contributor to the dust component.
Then, the mass fraction of aluminum of the total mass-loss rate is 
the mass fraction of aluminum in Al$_2$O$_3$ (53\%) divided by the
gas-to-dust ratio. \citet{Knapp1985} gave a typical gas-to-dust
ratio of 160 for oxygen-rich Mira variables. \citet{Norris2012}
give a larger gas-to-dust ratio of $\sim$600 for W\,Hya.
With these values, we obtain an aluminum mass fraction at the
outer shell radius $f_\mathrm{Al}(>R_\mathrm{out})$ between
$3.3\,\times\,10^{-3}$ and $8.8\,\times\,10^{-4}$, i.e. larger than 
$f_\mathrm{Al}(<R_\mathrm{in})$ by a factor between 60 and 16.
This means that the abundance of aluminum is too small
to explain typical observed gas-to-dust ratios with an 
Al$_2$O$_3$ dust shell alone.
As a result, it is likely that additional dust species
contribute to the dust shells of these types of Mira variables.
Because the Al$_2$O$_3$ shell is successful to describe
both the IRAS spectra as well as the MIDI photometry and visibility
of these sources, the additional species should have similar properties
as Al$_2$O$_3$,
in particular preserve the broad feature at 9--15\,$\mu$m
that is attributed to Al$_2$O$_3$ and not add additional features
that are not observed in the spectra of these stars, such as
the 'silicate features' near 9.7\,$\mu$m and 18\,$\mu$m.
We have shown above that the number density of aluminum can be high
enough in an extended atmosphere to match the required number density
of Al$_2$O$_3$ dust up to about 4 stellar radii, so that Al$_2$O$_3$
grains can be seed particles for additional dust condensation at larger
radii.
Candidates for such grains or inclusions are Fe (metallic iron) or 
(Fe,Mg)O (magnesiow\"{u}stite), 
\citep[cf. e.g.,][]{Ferrarotti2006, McDonald2010,Posch2002}.
Another possibility may be scattering iron-free silicates
(Mg$_2$SiO$_4$, forsterite), which may also not have a significant
silicate signature in the mid-IR if distributed in certain geometries,
such as a geometrically thin shell \citep{Ireland2005}.
\citet{Sacuto2013} could explain VLTI/MIDI spectra of RT~Vir, which
have a similar shape as those of S~Ori and R~Cnc, by a combination
of forsterite grains and Al$_2$O$_3$ grains.

\subsection{Coordinated SiO maser observations}
\label{sec:siomaser}
%MWI Double-check with Dave if this paragraph on the coordinated
%SiO maser observations is OK.
The VLTI/MIDI observations of our project were coordinated
with VLBA observations of the $v=1$, $J=1-0$ (43.1 GHz) and
$v=2$, $J=1-0$ (42.8 GHz) SiO maser emission toward the same
sources. \citet{Wittkowski2007} discussed the relationships between 
the photosphere, molecular layer, dust shell, and SiO maser shell 
at 4 epochs of S~Ori. The SiO maser ring radii were found to be
between 1.9 and 2.4 photospheric radii and to be co-located
with the extended molecular layers and the inner Al$_2$O$_3$
dust shell radii. Unpublished SiO maser ring radii for
GX~Mon, RR~Aql, and R~Cnc (Boboltz et al, in prep.) are on average
approximately 9.6\,mas, 8.8\,mas, and 12.0\,mas, respectively,
putting them as well between 1.9 and 2.3 photospheric radii using
the average photospheric angular diameters of Tab.~\ref{tab:par_avg_gxmon}.
These SiO maser ring radii of about 2 photospheric radii, possibly 
related to a shock front position, are consistent with other independent
measurements of Mira variables \citep{Cotton2010} and with theoretical
models \citep{Gray2009}. The co-location of our inner Al$_2$O$_3$ dust
shell radii between 1.9 and 2.2 photospheric radii 
(Tab.~\ref{tab:par_avg_gxmon}) with the SiO maser ring radii 
between 1.9 and 2.3 photospheric radii suggests that at these
radii silicon is mostly available as SiO in the gas phase, and that 
the dust grains at these small inner dust radii are thus unlikely to 
be related to SiO nucleation or to silicate dust.
If for some reason such dust could form at these radii despite the high temperature, 
the dust would deplete Si so that there would not be enough SiO gas left for the maser emission. 
Our following paper will be dedicated to a more detailed study on this topic.

\subsection{Effect of small amounts if silicate grains added to
an Al$_2$O$_3$ dust shell}
\label{sec:syntheticspectra}

\begin{figure}
\includegraphics[height=0.22\textheight]{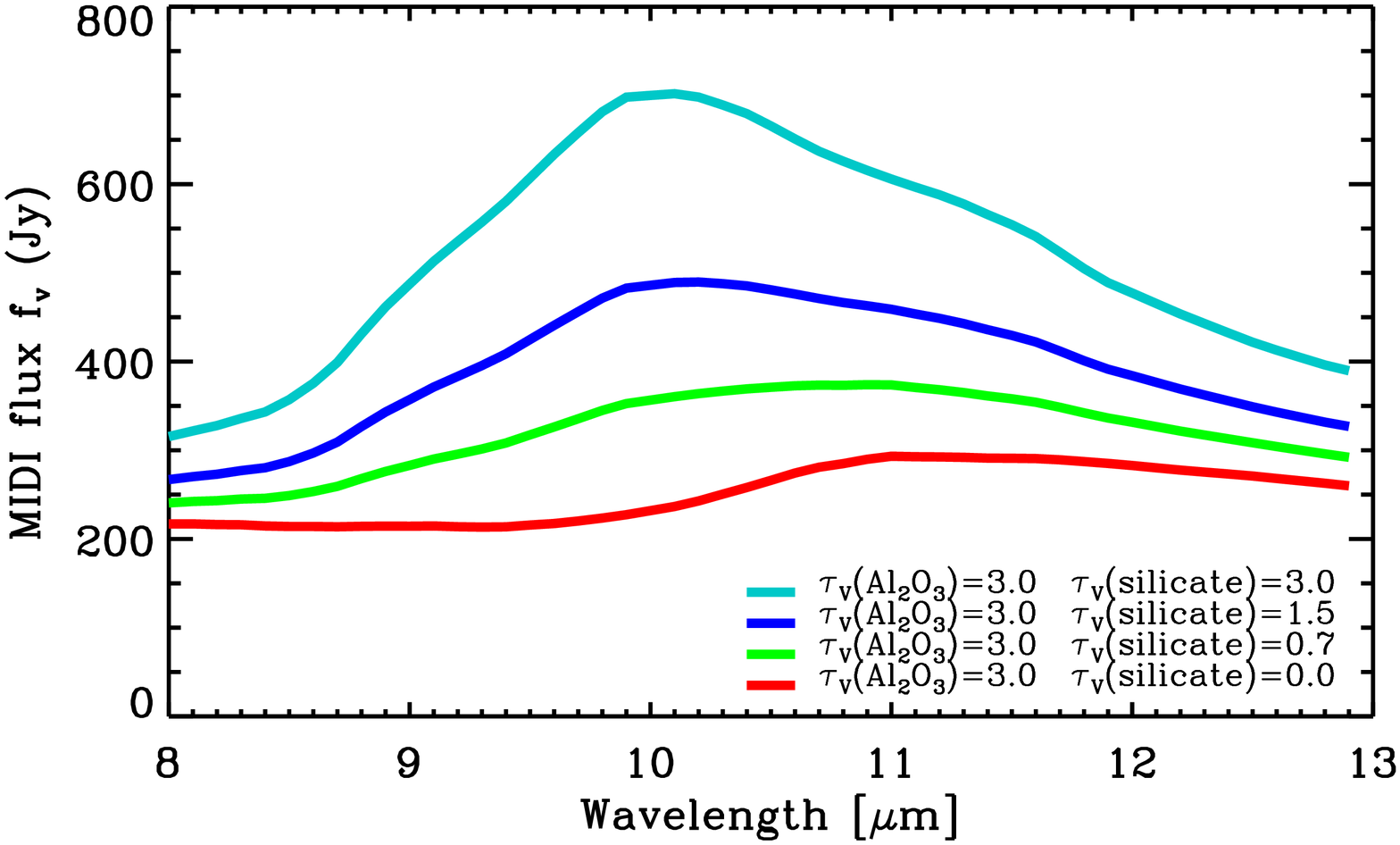}
\includegraphics[height=0.22\textheight]{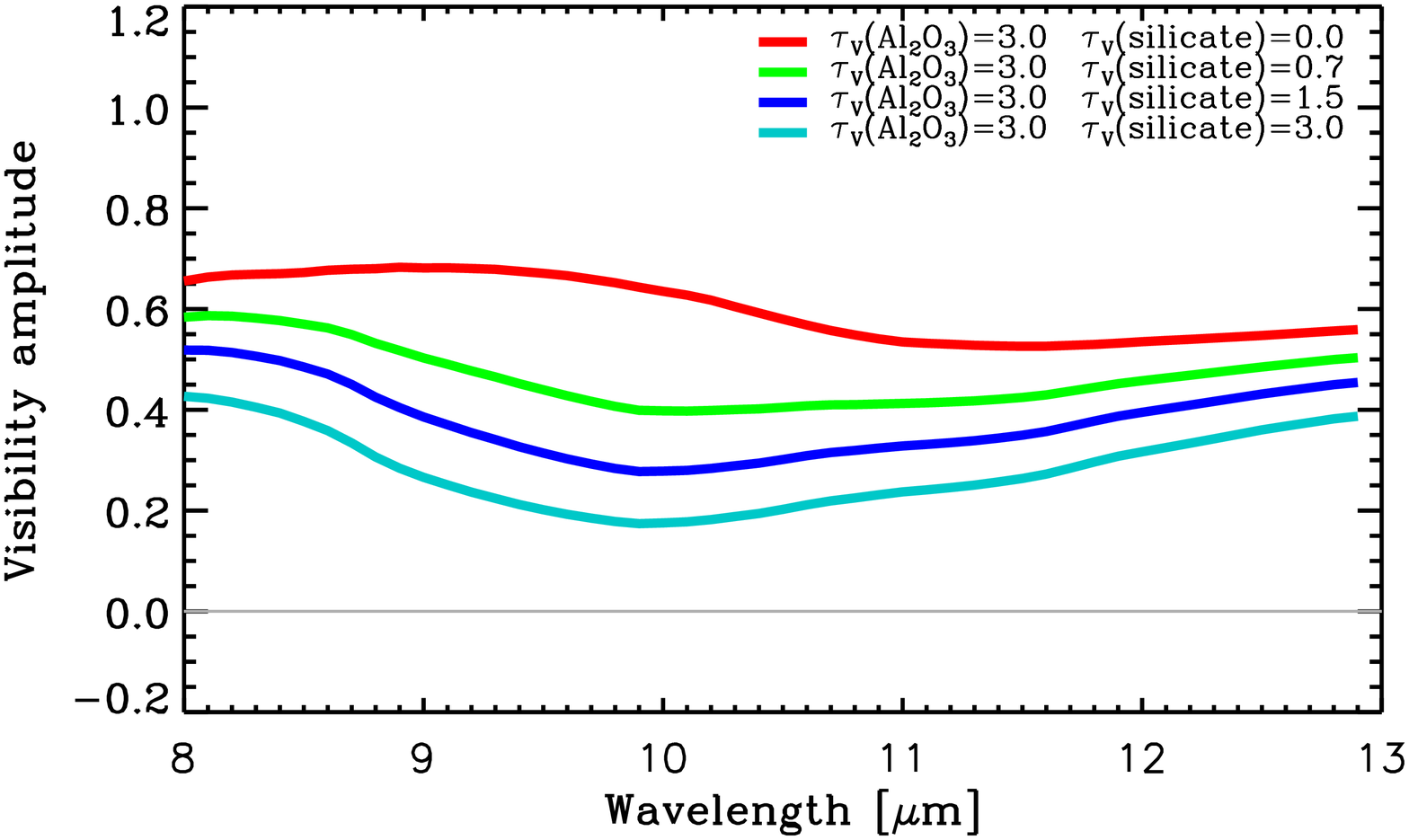}
\caption{
 Synthetic flux spectra (top) and visibility amplitude (bottom)
 based on the model parameters
 from Sec.~\ref{sec:bestfit}
 The simulation compares four different sets of dust 
 shell parameters where only the optical depth of the silicate 
 dust shell varies. The optical depth of an Al$_2$O$_3$ dust shell 
 $\tau_\mathrm{V}$ (Al$_2$O$_3$) is set to 3.0 and the optical depth 
 of a silicate dust shell $\tau_\mathrm{V}$ (silicate)
 is set to 0.0 (red), 0.7 (green), 1.5 (blue) and 3.0 (light blue).
 The full set of parameters can be found in the text.}
\label{testgxmon3}
\end{figure}

\citet[][their simulation 4]{Karovicova2011} demonstrated that
the addition of small amounts (10\%) of amorphous Al$_2$O$_3$ grains to a 
silicate dust shell are not detectable within the uncertainties
of our MIDI visibility and photometry measurements (differences $<$2\%). 
Here, we
investigate the opposite case, whether the addition of warm silicate
grains to an Al$_2$O$_3$ dust shell significantly
affects the shape and features of the mid-infrared
 visibility and photometry spectra.

For a given inner radius and density distribution, the amount of dust
grains of each compound was characterized by the optical depths at a
wavelength of 0.55 $\mu$m. 
For our simulation we used typical model parameters from 
Sec.\ref{sec:bestfit}. We used model M21n at a post-maximum
phase (0.1). The inner boundary radii were set to 
2.4 (Al$_2$O$_3$), and 4.0 (silicate), the power law index to 
3.0 (Al$_2$O$_3$) and 2.5 (silicate), the angular photospheric
diameter to 9.4, and the projected baseline length to 35m. We
considered models with the following combinations of optical depths:

\begin{itemize}
\item $\tau_\mathrm{V}$ (Al$_2$O$_3$)\,=\,3.0 and $\tau_\mathrm{V}$ (silicate)\,=\,0.0;
\item $\tau_\mathrm{V}$ (Al$_2$O$_3$)\,=\,3.0 and $\tau_\mathrm{V}$ (silicate)\,=\,0.7;
\item $\tau_\mathrm{V}$ (Al$_2$O$_3$)\,=\,3.0 and $\tau_\mathrm{V}$ (silicate)\,=\,1.5;
\item $\tau_\mathrm{V}$ (Al$_2$O$_3$)\,=\,3.0 and $\tau_\mathrm{V}$ (silicate)\,=\,3.0;
\end{itemize}

Figure~\ref{testgxmon3} shows the synthetic visibility and flux
spectra based on the model parameters above. 
The addition of a silicate dust shell with $\tau_\mathrm{V}$ (silicate)\,=\,0.7 leads to 
an increase of the flux at the wavelength of the silicate feature at 
9.8\,$\mu$ by 68\%, and to a decrease of the visibility by 75\%. These  
differences are significant compared to our observational uncertainties 
of 30-50\% for the flux and of 5-20\% for the visibilities.
We conclude that the addition of a silicate dust shell with
fairly low optical depth ($\tau_\mathrm{V}$ (silicate)\,=\,0.7 at a
wavelength of 0.55 $\mu$m) to the Al$_2$O$_3$ dust shell strongly affects
the shape and features of the photometry and visibility spectra.

\subsection{Variability}
\label{sec:variability}

\begin{figure}
\centering
\includegraphics[height=0.35\textheight,angle=90]{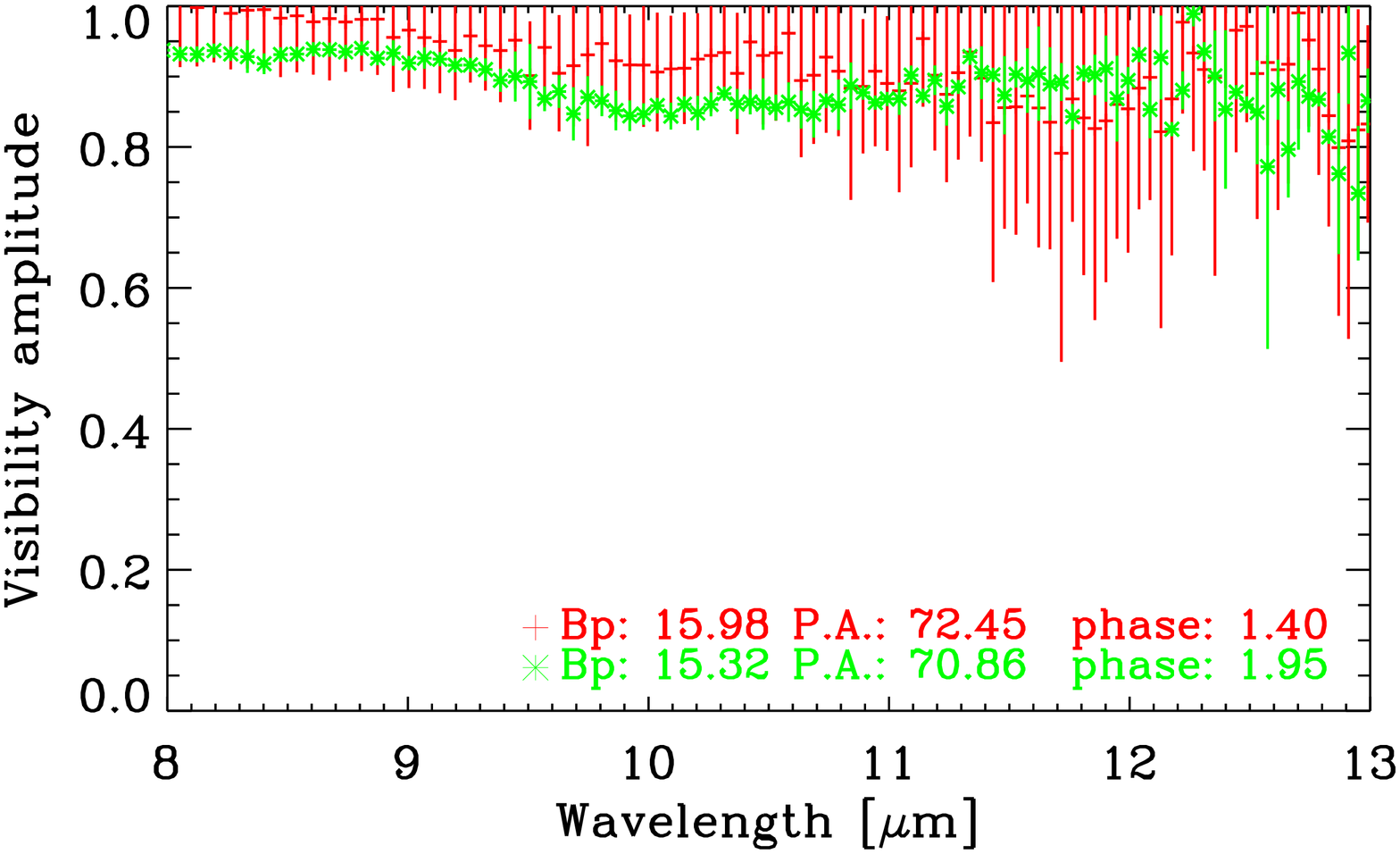}

\includegraphics[height=0.35\textheight,angle=90]{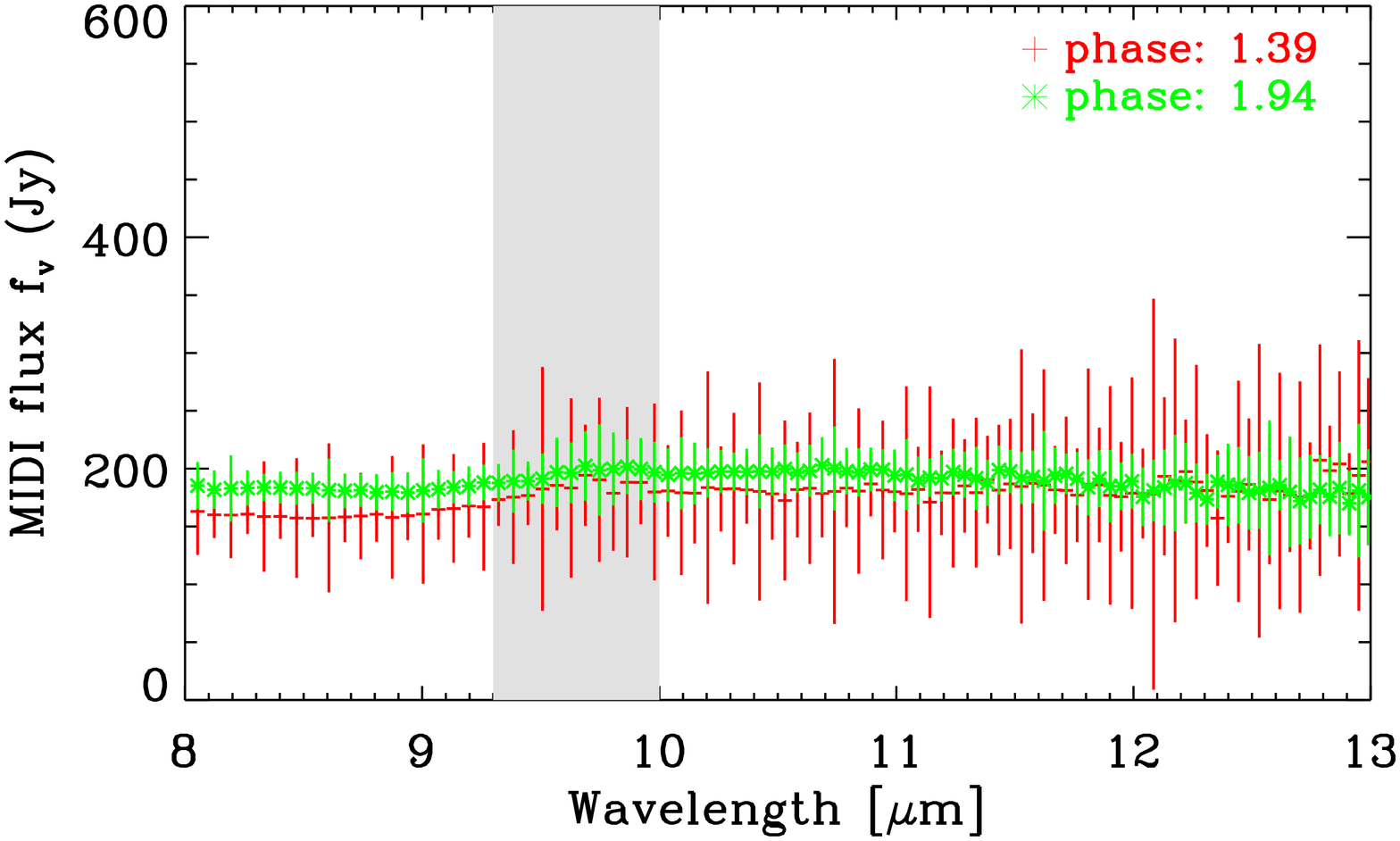}
\caption{Example of a test for intra-cycle visibility (top) and
photometry (bottom) variability of S~Ori.
The red lines denote a pre-minimum phase of 1.39 and the green lines 
a pre-maximum phase of 1.94. The visibility observations were obtained
at similar projected baseline lengths of $\sim$ 15\,m and similar
position angles of $\sim$71~$\degr$.
Both lines were computed as an average of data obtained at the
respective phase ($\pm$ 0.15), and for the visibility data
at the respective projected baseline length
($B_\mathrm{p}$\,$\pm$\,10\%) and position angle (PA\,$\pm$\,10\%) (see Tab.\ref{tab:variability_sori}).}
\label{fig:ic_monitoring}
\end{figure}

\begin{figure}
\centering
\includegraphics[height=0.35\textheight,angle=90]{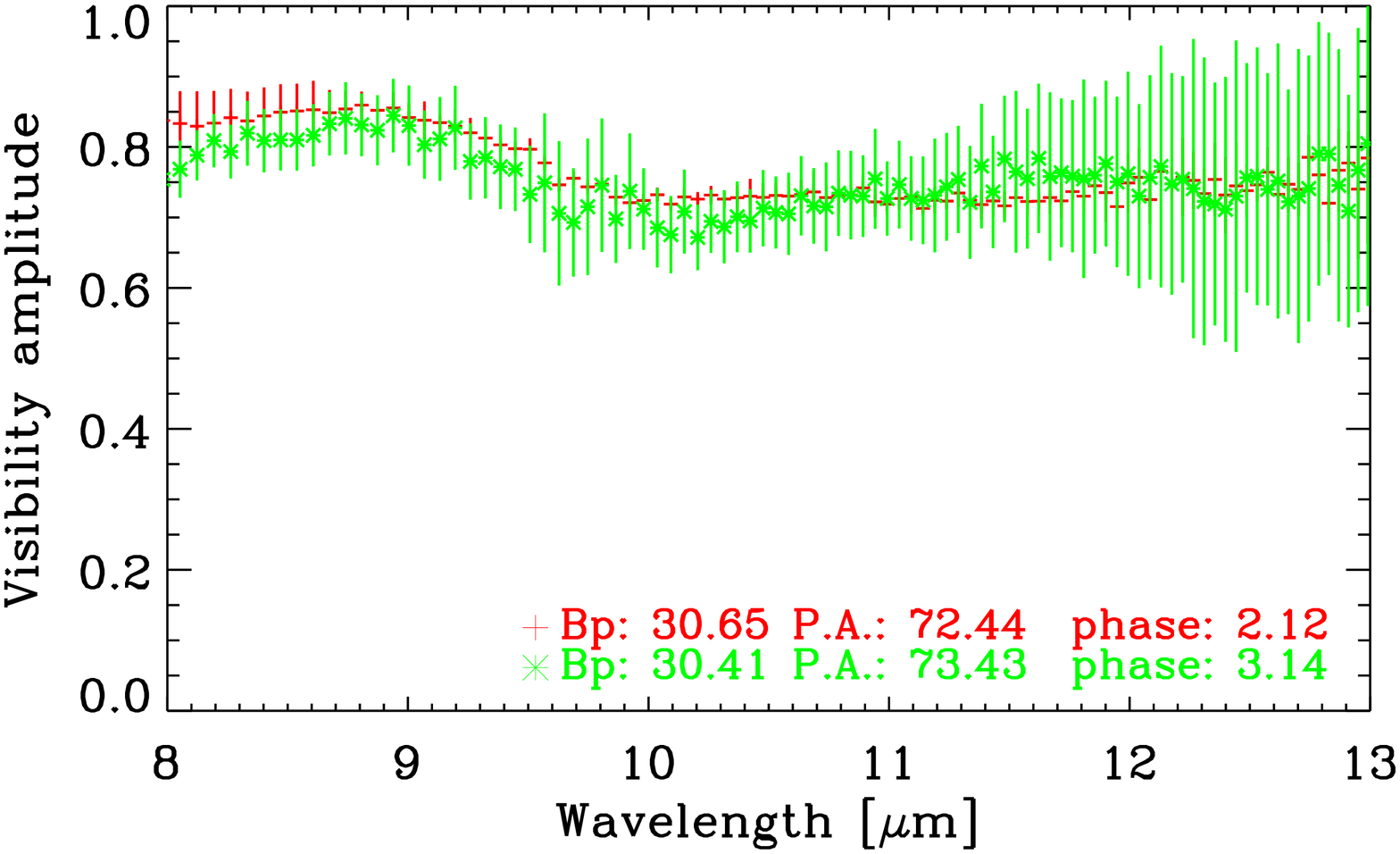}

\includegraphics[height=0.35\textheight,angle=90]{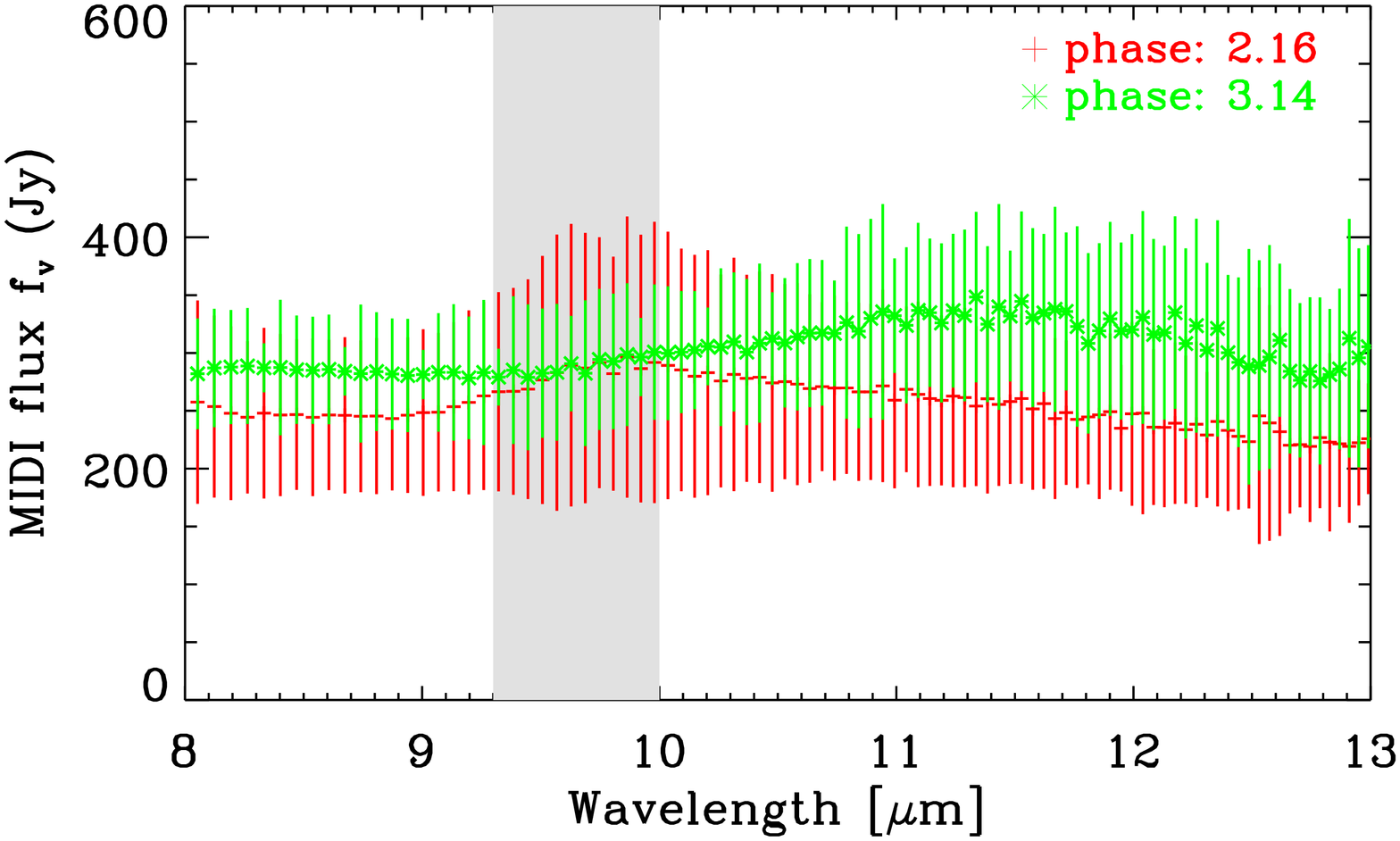}
\caption{Example of a test for cycle-to-cycle visibility (top)
and photometry (bottom) variability of S~Ori. The red lines denote a post-maximum
phase of 2.12 and the green lines denote a similar phase of 3.14 of the 
next cycle. The visibility observations were obtained at a similar
projected baseline length of $\sim$\,30\,m and at a similar position 
angle of $\sim$73\,$\degr$ (see Tab.\ref{tab:variability_sori}).}
\label{fig:ctc_monitoring}
\end{figure}

\onltab{8}{
\begin{table*}
\caption{Details of our tests to search for variability of S~Ori}
\centering
\begin{tabular}{l l l l l r l l}
\hline\hline
Star   & Vis/ & CTC/ & test& $B_\mathrm{p}$ & P.A. & ${\Phi_\mathrm{Vis}}$   & [DDMMYYYY]  \\
       &  Phot        &   IC         &                       &   [m]           &       [$\degr$]                     &           \\\hline

S~Ori  &     Vis     & CTC      & 1    & 12.77      &  71.93        & 1.38                   &  16022006.1  \\  %\cline{3-4}
       &             &          &      & 12.30      &  71.48        & 1.29                   &  12032007.1    \\
       &             &          &   2  & 61.74      &  70.88        & 1.93                   &  21092006.1, 17102006.1, 17102006.2     \\
       &             &          &      & 61.60      &  71.90        & 2.94                   &  02122007.1, 29122007.1     \\
       &             &          &    3 & 56.68      &  73.01        & 2.17                   &  20012007.1, 21012007.1    \\
       &             &          &      & 59.91      &  73.30        & 3.13                   &  06032008.1, 06032008.2, 06032008.3     \\
       &             &          &   4  & 30.65      &  72.44        & 2.12                   &  17122006.2, 21122006.1, 19012007.1      \\
       &             &          &      & 30.41      &  73.43        & 3.14                   &  13032008.1     \\
       &             &          &   5  & 26.24      &  72.28        & 3.19                   &  17122006.1     \\
       &             &          &      & 25.94      &  64.85        & 2.09                   &  01042008.1    \\
       &             &          &    6 & 47.78      &  72.78        & 3.88                   &  31122008.1, 31122008.2, 04032009.1     \\
       &             &          &      & 50.63      &  64.00        & 2.92                   &  10122007.1     \\
       &             &  IC      &    7 & 15.98      &  72.45        & 1.40                   &  22022006.1     \\
       &             &          &      & 15.32      &  70.36        & 1.95                   &  18102006.1, 19102006.1     \\
       &             &          &    8 & 63.97      &  72.74        & 2.09                   &  19122006.2     \\
       &             &          &      & 63.00      &  72.79        & 2.26                   &  10022007.1, 14032007.1    \\
       &             &          &      & 61.60      &  71.90        & 2.94                   &  02122007.1, 29122007.1     \\
       &             &          &    9 & 45.13      &  59.80        & 2.09                   &  19122006.1     \\
       &             &          &      & 50.63      &  64.00        & 2.92                   &  10122007.1     \\
       &             &          &    10& 70.82      &  125.25       & 3.81                   &  25122008.1, 25122008.2     \\
       &             &          &      & 70.58      &  125.40       & 3.96                   &  26022009.1, 26022009.2     \\
       &   Phot      &  CTC     &    11&            &               & 1.94                   &  18102006.1, 19102006.1, 21092006.1, 17102006.1, 17102006.2,  20102006.1    \\
       &             &          &      &            &               & 2.94                   &  02122007.1, 29122007.1   \\
       &             &          &    12&            &               & 3.96                   &  26022009.1, 26022009.2  \\
       &             &          &      &            &               & 2.16                   &  11012007.1, 11012007.2, 13012007.1, 17012007.1, 20012007.1, \\
       &             &          &    13&            &               &                        &  21012007.1, 19012007.1, 18012007.1 \\
       &             &          &      &            &               & 3.14                   &  06032008.1, 06032008.2, 06032008.3, 13032008.1, 01042008.1   \\\
       &             &          &      &            &               & 2.26                   &  12032007.2, 11022007.1, 12032007.1, 10022007.1, 14032007.1, 12022007.1   \\
       &             &          &      &            &               & 1.39                   &  16022006.1, 22022006.1   \\
       &             &  IC      &    14&            &               & 1.39                   &  16022006.1, 22022006.1   \\
       &             &          &      &            &               & 1.94                   &  18102006.1, 19102006.1, 21092006.1, 17102006.1, 17102006.2, 20102006.1  \\
       &             &          &    15&            &               & 2.26                   &  12032007.2, 11022007.1, 12032007.1, 10022007.1, 14032007.1, 12022007.1   \\
       &             &          &      &            &               & 2.16                   &  11012007.1, 11012007.2, 13012007.1, 17012007.1, 20012007.1, 21012007.1, \\
       &             &          &      &            &               &                        &  19012007.1, 18012007.1 \\
       &             &          &      &            &               & 2.08                   &  19122006.2, 14112006.1, 17122006.2, 21122006.1, 17122006.1, 19122006.1  \\
       &             &          &      &            &               & 2.94                   &  02122007.1, 29122007.1   \\
       &             &          &      &            &               & 3.14                   &  06032008.1, 06032008.2, 06032008.3, 13032008.1, 01042008.1   \\
       &             &          &   16 &            &               & 3.82                   &  31122008.1, 31122008.2, 25122008.1, 25122008.2   \\
       &             &          &      &            &               & 3.96                   &  26022009.1, 26022009.2   \\\hline
\end{tabular}
\tablefoot{The table lists the star, the type of measurement: 
interferometric (Vis) or photometric (Phot), the type of examined variations: intra-cycle (IC) 
or cycle-to-cycle (CTC), the test, the average projected baseline [m], the
the averaged position angle [$\degr$], the averaged pulsation phase, and the
dates of observations. The number after the point indicates the science object if it was observed multiple times over the night}
\label{tab:variability_sori}
\end{table*}
}

\onltab{9}{
\begin{table*}
\caption{Details of our tests to search for variability of GX~Mon}
\centering
\begin{tabular}{l l l l l r l l}
\hline\hline
Star   & Vis/ & CTC/ & test& $B_\mathrm{p}$ & P.A. & Epoch   & [DDMMYYYY]  \\
       &  Phot        &   IC         &                       &   [m]           &       [$\deg$]                     &           \\\hline

GX~Mon  &     Vis     & CTC    &      1& 14.62     &  66.85        & J/K  & 14032008.1, 28032008.1   \\  %\cline{3-4}
        &             &        &       & 13.51     &  74.92        & A    &  18102006.1    \\
        &             &        &      2& 62.74     &  73.99        & C/D  & 11112006.1, 20122006.1 \\
        &             &        &       & 62.40     &  70.21        & I    & 22022008.1, 06032008.4, 06032008.5\\
        &             &        &      3& 30.23     &  72.96        & D    & 16122006.1, 17122006.1\\
        &             &        &       & 28.85     &  66.29        & K    & 01042008.2, 01042008.3, 01042008.4\\
        &             &  IC    &      4& 15.60     &  70.07        & B    & 18032006.2\\
        &             &        &       & 15.40     &  73.66        & D/E  & 14122006.2, 11012007.1, 11012007.2, 13012007.1, 11022007.1, 11022007.2\\
        &             &        &      5& 62.74     &  73.99        & C/D  & 11112006.1, 20122006.1\\
        &             &        &       & 61.14     &  73.53        & F    & 10022007.1, 10022007.2\\
        &             &        &      6& 30.23     &  72.96        & D    & 16122006.1, 17122006.1\\
        &             &        &       & 29.60     &  72.82        & F    & 21012007.1, 21012007.2, 12022007.1\\
        &     Phot    &  CTC   &      7&           &               & C    & 18102006.1\\
        &             &        &       &           &               & I/J/K& 14032008.1, 28032008.1, 22022008.1, 06032008.4, 06032008.5, \\
        &             &        &       &           &               &      & 13032008.2, 01042008.2, 01042008.3, 01042008.4\\
        &             &  IC    &      8&           &               & B    & 18032006.2\\
        &             &        &       &           &               & C/D  & 14122006.2, 18102006.1, 11112006.1, 20122006.1, 16122006.1, 17122006.1 \\
        &             &        &       &           &               & F    & 11022007.1, 11022007.2, 10022007.1, 10022007.2, 09022007.1,\\
        &             &        &       &           &               &      & 09022007.2, 12022007.1\\
        &             &        &      9&           &               & I/J/K& 14032008.1, 28032008.1, 22022008.1, 06032008.4, 06032008.5, \\
        &             &        &       &           &               &      & 13032008.2, 01042008.2, 01042008.3, 01042008.4\\
        &             &        &       &           &               & H    & 10012008.1, 12012008.1, 29122007.1, 13012008.1, 10122007.1, 11012008.1\\\hline
\end{tabular}
\tablefoot{The phases are uncertain, and therefore they are omitted and replaced by epochs.}
\label{tab:variability_gxmon}
\end{table*}
}

We investigated whether our visibility and photometry data exhibit
a sign of intra-cycle or cycle-to-cycle variability. Such a variability
could be caused by dust formation that occurs preferentially at 
certain variability phases or only during some cycles.
Variability at mid-infrared wavelengths has previously been noted by 
\citet{Lopez1997}, \citet{Tevousjan2004},
\citet{Ohnaka2007}, \citet{Zhao-Geisler2012}.

Our observations of S~Ori included 13 epochs of several phases between 
a pre-maximum phase of 0.9 and a pre-minimum phase of 0.4 
(see Sect.~\ref{sec:observations}), and some of these phases were
obtained at several cycles. Likewise, the GX~Mon observations were
obtained at 11 different epochs, but for which we could not assign certain
phases to the epochs because of the unknown lightcurve of GX~Mon.
The R~Cnc observations were obtained at only 2 epochs, both at 
near-maximum phases, so that R~Cnc could not be probed for variability.

We directly compared visibility data that were obtained at similar projected
baseline length and position angles, which is important for a meaningful
comparison. Visibility observations obtained 
at different baseline lengths show different values, so that a direct
comparison is not possible. Visibility observations obtained at the
same baseline length but at different position angles could also be
caused by an asymmetric intensity distribution, which could be confused
with variability. Our direct intra-cycle and cycle-to-cycle comparisons
follow the investigations by \citet{Karovicova2011} of RR~Aql.
Tabs.~\ref{tab:variability_sori}--\ref{tab:variability_gxmon} 
detail for our comparisons of S~Ori and
GX~Mon the combinations of projected baseline
lengths, position angles, and variability phases.
Our direct visibility comparisons did not exhibit any significant 
variability for S~Ori and GX~Mon within our visibility accuracies 
of $\sim$ 5--20\%, and within the limits of the phase coverage and the
available baseline lengths.
Our MIDI photometry data showed large uncertainties at some epochs of
up to 50\%. The data of GX~Mon showed
a variability, both intra-cycle and cycle-to-cycle, 
at the level of $\sim$\,2-3\,$\sigma$ (test 7 \& 9 of Tab.~\ref{tab:variability_gxmon}), which was most
pronounced near the silicate emission feature at 9.8\,$\mu$m.
However, because of the large uncertainties of our MIDI photometry,
this variability is not conclusive, and follow-up observations
with a higher photometric accuracy, for example employing the
instrument VISIR at the VLT, are needed to confirm a photometric
variability.
As an illustration, Fig.~\ref{fig:ic_monitoring} shows an example
of a test for intra-cycle variability of S~Ori (phases 0.4 and 0.9,
observed with a projected baseline length of $\sim$\,15\,m at position
angle $\sim$71\,$\degr$.), and Fig.~\ref{fig:ctc_monitoring} shows
an example of a test for cycle-to-cycle variability of S~Ori 
(phases $\sim$\,0.1 observed at a baseline length of $\sim$\,30\,m
and position angle $\sim$73\,$\degr$.).

It has to be noted that our direct comparisons of the visibility and 
photometry spectra had some limitations. As mentioned above, the 
phase coverage of our observations was not complete. Most observations
of S~Ori and all observations of R~Cnc were obtained at near-maximum
phases, and we could not assign variability phases to the epochs
of GX~Mon. In addition, \citet{Karovicova2011} showed that the expected
variability is small at mid-infrared wavelengths, and is difficult to 
detect within the uncertainties of the MIDI data. In addition, they showed
that differences of the visibility spectra can best be detected using 
certain baseline lengths so that the dust shells are neither unresolved
nor over-resolved. 
As a result, we can not exclude that intra-cycle or cycle-to-cycle 
variability of our sources could be detected with a more complete coverage
of phases and baseline lengths or with a higher precision.

\subsection{Asymmetries}

\begin{figure}
\includegraphics[height=0.37\textheight,angle=90]{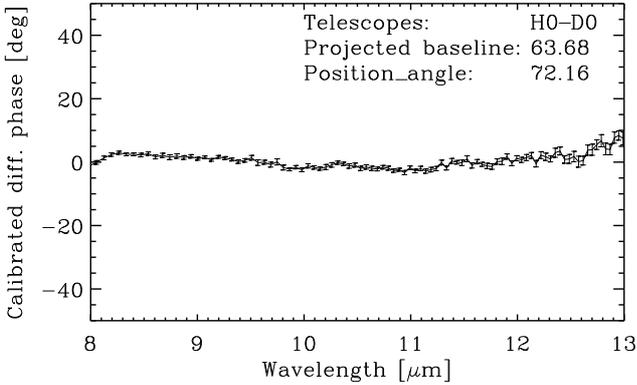}
\caption[]{Calibrated differential phase [$\degr$] of one of the observations 
of S~Ori obtained on 10/02/2007 with a projected baseline length 
of $\sim$\,64\,m.}
%Iva: I have many flatter examples, but I decided to show this one
\label{fig:sori_diff}
\end{figure}

Asymmetric intensity distributions can be probed in two different
ways using VLTI/MIDI data. Asymmetries can be investigated 
either by comparing visibility data 
obtained at the same projected baseline length, but different position 
angles, or by analyzing the differential phases that are computed
as a part of the EWS data reduction. 
The latter method was previously
used by \citet{Ohnaka2008,Paladini2012,Sacuto2013} to detect
asymmetric intensity distributions of the environments of AGB stars.

Our data of S~Ori were mostly obtained at position angles around
70\,$\degr$. Only 4 observations were obtained at a different position
angle of $\sim$\,125\,$\degr$, but also at a different baseline length
(different by at least 10\,m). Our GX~Mon observations were all obtained
at similar position angles between $\sim$\,60\,$\degr$ and $\sim$\,80\,$\degr$.
Our R~Cnc observations were obtained at two different position angles,
but these also correspond to clearly different baseline lengths.
This means that we could not probe our sources for asymmetries
using a direct comparison of visibility amplitudes.

The differential phases show deviations from zero by less then
5\,$\degr$ for S~Ori and R~Cnc, and less than 10\,$\degr$ for GX~Mon.
These values are of the order of typical instrumental and atmospheric
differential phases \citep{Ohnaka2008}. This means that
our data do not
show an indication of asymmetric intensity distributions.
Fig.~\ref{fig:sori_diff} shows one example of a differential phase of S~Ori,
obtained with one of the longest available projected baseline lengths
of $\sim$\,64\,m.
The differential phase probes scales corresponding to the 
projected baseline length. We can not exclude that asymmetries of our
sources might be detectable with longer projected baseline lengths. 

\section{Discussion of the dust condensation process}
\label{sec:dust}

In our project we focused on an investigation of the
dust condensation process based on spatially and spectrally
resolved mid-infrared interferometry. In the following
we discuss our results in the context of the scenarios on
the formation of seed particles and on the dust condensation sequence that
have been discussed in the literature.

\subsection{Seed particles}
It is known that the candidates for the first kind of
grains forming from the gas phase in the environment of oxygen rich
stars should condense out of abundant
species, the reactions should not be blocked by high bond energies,
and the condensation temperature should be high enough to allow the dust to
survive in the hostile environment \citep{Gail1999}. 
The dust temperatures at the inner edge of dust shells were observed
to be at least around 1000\,K \citep{Gail2013}. However, the smallest
observed dust radii of 1.5--3 stellar radii 
\citep{Ireland2005,Wittkowski2007,Norris2012}
correspond to higher dust temperatures of 1300--1400\,K.
Generally, three options for the condensation of seed particles are being
discussed in the recent literature.

(I) SiO is one of the most abundant species, but its efficient condensation 
takes place only below 600 K. However, new measurements seem to indicate 
higher SiO condensation temperatures, so that silicate dust formation triggered
by cluster formation of SiO may be compatible with observed dust temperatures
around 1000\,K \citep{Gail2013}. 

(II)
Magnesium oxide can be ruled out as seed particle due to its low
stability. All other species are less abundant. According 
to \citet{Gail1999} aluminum oxides and titanium oxides remain as 
potentially possible primary condensing species. 
A possibility is that dust 
formation starts with the formation of TiO clusters, which serve as 
growth centers both for aluminum oxides and silicates. However, aluminum 
can also condense in circumstellar shells of its own due to the temperature 
gap between the formation of TiO seed nuclei and the growth of silicate 
dust \citep{Gail1999}.
The high condensation temperature of aluminum oxide (Al$_2$O$_3$)
of $\sim$1400\,K in circumstellar environments allows these refractory 
grains to exist relatively close to the star. In addition, 
Al$_2$O$_3$ grains can condense before silicate grains, not
only because of the higher condensation temperature, but also due to
the greater affinity of oxygen for aluminum compared to silicon
\citep{Stencel1990}, together making it a good candidate as a seed particle.
Note that \citet{Gail2013} also discussed that Al$_2$O$_3$ grains and SiO 
clustering are mutually exclusive as seed particles.
Many studies suggested that Al$_2$O$_3$ grains can become coated
with silicates at larger radii where the dust temperature is lower
and can serve as seed nuclei for the
subsequent silicate formation \citep[e.g.:][]{Deguchi1980, Vardya1986,
Kozasa1997b, Kozasa1997a}. 
Aluminum dust as well as for instance alloys of iron and nickel
are difficult to detect because of the lack of diagnostically useful 
emission bands \citep{McDonald2010}. 
Amorphous Al$_2$O$_3$ is associated to a broad spectral feature at wavelengths
between 9\,$\mu$m to 15\,$\mu$m \citep{Koike1995,Begemann1997} due to a
broad Al-O vibrational band.
The origin of the 13\,$\mu$m feature is still under debate \citep{DePew2006}.
\citet{Vardya1986, ClaytonNittler2004} assign it to crystalline Al$_2$O$_3$. 
On the other hand, \citet{Speck1998} attribute it to silica SiO$_2$, and \citet{Posch1999, Smolders2012} to spinel MgAl$_2$O$_4$.
A detailed investigation is still required. However, crystalline oxides are expected to be a minor component of most oxygen-rich dust shells.

(III) 
\citet{Ireland2005} and \citet{Norris2012} discussed that large 
($\sim$300\,nm) scattering iron-free, magnesium-rich silicate (forsterite) 
grains can be stable at small radii down to observed radii of 1.5 
stellar radii. Similarly, \citet{Hofner2008,Bladh2012,Bladh2013}
proposed micron-sized iron-free silicates as wind-driving grains.
However, both teams also mention Al$_2$O$_3$ grains as an alternative
\citep{Ireland2005,Norris2012} or additional \citep{Sacuto2013}
dust species.
\citet{Goumans2012} favor the formation of magnesium-rich iron-free silicates
via heteromolecular condensation based on Mg, SiO, and H$_2$O.

The small observed dust radii of 1.9--2.2 stellar radii based on our models
of S~Ori, R~Cnc, and GX~Mon, and the corresponding high dust temperatures
of 1200--1350\,K support a scenario where Al$_2$O$_3$ grains serve
as seed particles for the dust formation (scenario II) of these targets. 
This scenario is further supported
by the good agreement of the MIDI spectra and the models including 
Al$_2$O$_3$ grains, in particular regarding the broad spectral feature
at $\sim$9--15\,$\mu$m that is attributed to Al$_2$O$_3$.
In addition, we have shown that the aluminum number density in an 
extended stellar atmosphere can be sufficiently high to match the 
number densities of Al$_2$O$_3$ grains of our dust shell model between 
$\sim$ 2 and up to $\sim$ 4 stellar radii (Sect. \ref{sec:alabundance}).  
The co-location of SiO maser emission and the inner dust radii at about
2 stellar radii for the same sources (Sect.~\ref{sec:siomaser})
show that silicon is available 
as SiO in the gas phase at these radii, indicating that SiO clustering 
(scenario I) and the formation of silicates at these small radii 
(scenario III) are unlikely.
It is not clear if Al$_2$O$_3$ grains also play a role in the case of RR~Aql,
where our best-fit dust model consisted only of silicate grains with a larger
inner radius of about 4 stellar radii \citep{Karovicova2011}. 
It is possible that also in this case Al$_2$O$_3$ grains serve
as seed particles at smaller inner radii, but that they are not visible behind 
the silicate grains in the model comparison. Indeed, model simulations 
have shown that the 8--13\,$\mu$m visibility and flux values are not
sensitive to the addition of an Al$_2$O$_3$ dust shell with low
optical depth (10--20\%) to a silicate dust shell within the VLTI/MIDI
uncertainties \citep{Karovicova2011}. 
However, RR~Aql may also be an example of another route of direct
silicate dust formation at larger radii via scenarios I/III.

\subsection{Dust condensation sequence}
Our results were consistent with the approach by \citet{Little-Marenin1990}
to classify Mira variable AGB stars into three groups that can be described
by Al$_2$O$_3$ shells, silicate shells, or a mix thereof. 
Our targets S~Ori and R~Cnc could be well described by
Al$_2$O$_3$ shells with small inner radii of 1.9--2.2 photospheric radii.
GX~Mon could be well described with a mix of an Al$_2$O$_3$ shell with
an inner radius of 2.1 photospheric radii, and a silicate shell with
an inner radius of 4.6 photospheric radii. RR~Aql could be
well described by a silicate dust shell alone with an inner radius 
of 4.1 photospheric radii.

\citet{Little-Marenin1990} suggested that these three groups of
sources follow an
evolutionary sequence in the spectral features. This sequence starts
with a featureless continuum, then develops toward strong silicate
features. This is in accordance with a theoretical dust condensation
sequence by \citet{Lattimer1978} that predicts primary condensation of
refractory oxides. If in the cooler regions, further out in the
circumstellar shell, the densities are still high enough, the formation of
silicate dust can occur. In cases where the densities are not high
enough for the formation of the silicate dust, the dust condensation
sequence can ``freeze-out''. This would imply that stars with low
mass-loss rates will predominantly exhibit an oxide mineralogy while
stars with high mass-loss rates will form substantial amounts of
silicate dust. This scenario was confirmed e.g.by \citet{Stencel1990, Blommaert2006,Posch1999, Heras2005, Smolders2012, Zeidler2013}
showing that for low mass-loss rate stars the
dust content is dominated by Al$_2$O$_3$ grains.

A second scenario was suggested by \citet{Sloan1998}. According
to these authors, the chemistry of the dust (i.e. the condensation
sequence) is driven by the C/O abundance of the out-flowing gas. In
this scenario, less evolved stars - stars with low C/O ratio - produce
larger amounts of silicate grains compared to alumina grains due to
the abundance of oxygen atoms, and therefore we should see strong
silicate features. For more evolved stars - stars with C/O ratio close
to the unity - most of the oxygen will be bound up in CO molecules
and in-sufficient oxygen will remain for grain condensation in order to form
silicate grains. Al$_2$O$_3$ grains will therefore dominate in the
dust shell. The second scenario suggests an increasing oxide dust
content with an increasing mass-loss rate during the evolution of the
star along the AGB.

\begin{table*}
\caption{Dust condensation sequence}
\centering 
\begin{tabular}{ l l l l l r }
\hline\hline 
Star & Dust chemistry &$\tau_\mathrm{V}$(Al$_2$O$_3$) &$\tau_\mathrm{V}$(silicate) & T$_\mathrm{Phot}$ [K] & Mass-loss rate [M$_{\odot}$/year]\\\hline
R~Cnc & Al$_2$O$_3$&1.35$\pm$0.2 & 0.0 &  2604$\pm$300 &0.2\,$\times$\,10$^{-7}$ $^{a}$ \\
S~Ori & Al$_2$O$_3$ & 1.5$\pm$0.5 & 0.0 &  2627$\pm$300 &2.2\,$\times$\,10$^{-7}$ $^{a}$ \\
RR~Aql  & silicate & 0.0& 2.8$\pm$0.8&  2420$\pm$200   &9.1\,$\times$\,10$^{-7}$ $^{b}$\\
GX~Mon & Al$_2$O$_3$ and silicate  & 1.9$\pm$0.6 & 3.2$\pm$0.5 &  2173$\pm$300  & 54\,$\times$\,10$^{-7}$ $^{b}$\\
\hline
\end{tabular}
\tablefoot{The table lists the star, the dust chemistry, the optical depth $\tau_\mathrm{V}$, the photospheric temperature, and the mass-loss rate adopted from $^{a}$\citet{Young1995}, $^{b}$ \citet{Loup1993}.} 
\label{tab:massloss}
\end{table*}

Tab.~\ref{tab:massloss} lists for our targets
the dust chemistry and the optical depth together with the mass loss
rates adopted from the literature. 
These results point in the direction of the
hypothesis by \citet{Little-Marenin1990} and
\citet{Blommaert2006} that the dust content of stars with low
mass-loss rates is dominated by Al$_2$O$_3$ and other 
oxide grains while the dust
content of stars with higher mass-loss rates predominantly exhibits
substantial amounts of silicates.

However, as outlined in Sect.~\ref{sec:alabundance},
it is unlikely that in those sources that can be described by only
an Al$_2$O$_3$ shell (S~Ori and R~Cnc in our sample), Al$_2$O$_3$ grains 
can indeed alone explain
the observed gas-to-dust ratios of the stellar outflows, owing to the 
relatively low abundance of aluminum. This means that it is unlikely
that the dust condensation sequence already freezes out directly after
the formation of Al$_2$O$_3$ grains.
It is more likely that Al$_2$O$_3$ 
grains form at small inner radii (about 2 stellar radii), see above,
and serve as seed particles for grains at larger radii (from about 4 stellar
radii). For those sources that do not show the typical silicate features
in the spectra and that can be well described by Al$_2$O$_3$ shells alone,
the additional dust species need to have properties that
preserve the spectral signature of Al$_2$O$_3$, in particular
the broad feature at $\sim$9--15\,$\mu$m, and that do not add the typical
silicate features at 9.7\,$\mu$m and 18\,$\mu$m, which is not seen
in the spectra of these sources. 
As shown in Sect. ~\ref{sec:syntheticspectra},
already a small amount of additional warm silicate grains would 
significantly affect the shape and features of the visibility and 
photometry spectra.

The class of sources that can be well described by a combination of
an Al$_2$O$_3$ shell with an inner radius around 2 photospheric radii
and a silicate shell with an inner radius around 4--5 photospheric radii 
(GX~Mon in our sample), as well as those that can be described by
a silicate shell alone (RR~Aql in our sample) show the typical silicate 
emission features at roughly 9.7 $\mu$m and 18 $\mu$m.
These sources also show larger mass-loss rates and larger overall
optical depths of the dust shell (cf. Tab.~\ref{tab:massloss}).
For these sources, the dust condensation sequence continues beyond
that described above.
The silicate emission features
are generally believed to be associated with silicate condensates or
more precisely to amorphous silicates known as glasses. The features are
associated with vibrational modes of the fundamental SiO$_4$
tetrahedron in silicate condensates of a largely disordered lattice
structure \citep{Gail1999}. The main dust
components responsible for most of the dust opacity and the emission
of infrared radiation are olivine and pyroxene \citep{Vollmer2007}, 
non-stoichiometric varieties, with random ratios between the Fe/Mg, Si, 
together with oxygen. The condensation of amorphous silicates
proceeds at temperatures below the glass temperature. An amorphous
structure is formed by molecules that stick to the surface of the
grain, and immediately freeze out without having sufficient energy to
find energetically more favorable lattice positions. The spectrometers on
board of ISO showed that in addition to amorphous silicates also the
crystalline silicates were found in the circumstellar environment of
evolved star \citep{Waters1996}. It is under debate how both
crystalline and amorphous silicates can be formed in the outflow of
evolved stars. It is believed that crystalline silicates are not
present in the dust shells around low mass-loss rate AGB stars with
optically thin dust shells. However, \citet{Kemper2001} showed that due
to a temperature difference between amorphous and crystalline
silicates it is possible to include up to 40\% of crystalline silicate
material in the circumstellar dust shell, without the spectra showing
the characteristic spectral features.

\section{Summary and conclusions}
We presented multi-epoch spectrally and spatially resolved
mid-infrared interferometric observations of the
oxygen-rich Mira variables S~Ori, GX~Mon, and R~Cnc obtained
with the MIDI instrument at the Very Large Telescope Interferometer (VLTI). 
This study represents a continuation of
previous results by \citet{Wittkowski2007} and
\citet{Karovicova2011}, who compared multi-epoch VLTI/MIDI data of 
the Mira variables S~Ori and RR~Aql to radiative transfer models
of the dust shell. 
The main goal of this project was to trace 
the dust formation process as a function of mass-loss rate, 
distance from the photosphere, and pulsation phase and cycle.

Our data did not show significant signs of variability or asymmetry
within their uncertainties, and within our limited baseline coverage and  
phase coverage.
Intra-cycle and cycle-to-cycle photometry variations may be present
at the level of $\sim$2-3$\sigma$ for GX~Mon.
However, because of the large uncertainties, this photometric
variability is not conclusive, and observations with a higher
photometric accuracy (e.g. VISIR at the VLT) are needed for 
a confirmation.
A variability of the visibility may be expected with mid-infrared
interferometric observations that cover more phases and use
optimized baseline lengths, such that the dust shells are neither
barely resolved nor over-resolved \citep[cf.][]{Karovicova2011}.

We modeled the observed data with an ad-hoc radiative transfer
model of the dust shell using the radiative transfer code mcsim\_mpi
\citep{Ohnaka2006}. 
We used a series of dust-free dynamic model
atmospheres based on self-excited pulsation models \citep[M
series,][]{Ireland2004b,Ireland2004a} to describe the intensity
profile of the central source. 
Following the successful description
of IRAS data of a number of Mira stars by \citet{Lorenz-Martins2000},
we used two dust shells that can have different parameters, 
one using amorphous Al$_2$O$_3$ grains \citep{Koike1995,Begemann1997} 
and one using warm 'astronomical' silicate grains \citep{Ossenkopf1992}.

We showed that the photometric and visibility spectra of all our sources
can be well described by our modeling approach. We found 
for all epochs the best-fit 
dust shell parameters including the optical depths of
the dust shells, the inner boundary radii, the power-law indexes of the
density distributions, as well as the photospheric angular diameter.

The parameters of our best-fit models were listed in 
Tab.~\ref{tab:par_avg_gxmon} and were described in Sect.~\ref{sec:bestfit}.

In summary, our Mira variable sources can be classified in three groups,
as suggested by \citet{Lorenz-Martins2000},
where the dust shell can be characterized by (I) an Al$_2$O$_3$ shell
alone (S~Ori and R~Cnc), (II) a silicate shell alone (RR~Aql), and
(III) a combination of an Al$_2$O$_3$ shell and a silicate shell (GX~Mon).
Hereby, the inner dust shell radii of Al$_2$O$_3$ are located closer to the 
star at distances of 1.9--2.2 photospheric radii, and the inner dust shell 
radii of the silicates are farther away at distances of 4--5 photospheric radii.
The best-fit photospheric radii are consistent with independent 
estimates. The derived absolute photospheric radii of 
$\sim$350--675\,R$_\odot$ and effective temperatures of $\sim$2200--2600\,K
are consistent with typical values of Mira variables.
The dust temperatures at the inner radii of 1200--1350\,K
(Al$_2$O$_3$) and $\sim$1100\,K (silicates) are consistent
with the condensation temperatures of Al$_2$O$_3$ and silicate grains,
respectively. This increases the confidence in our modeling approach.
The inner dust shell radii of the Al$_2$O$_3$ grains are co-located
with the extended atmospheres and with SiO maser emission for all
our sources.
Our results combined with mass loss rates adopted from the literature
support the hypothesis that stars with low mass-loss rates are those
that can be described with Al$_2$O$_3$ grains alone (group I),
and stars with higher mass-loss contain significant amounts of warm
silicates displaying the prominent silicate features (groups II and III).
It should be noted that even tough the results are very interesting,
there are, due to the rather small sample size (four stars), still relatively
far from reaching solid conclusions.

For group (I), sources which are characterized by only an Al$_2$O$_3$ shell,
we confirmed that the number density of aluminum in a sufficiently
extended atmosphere can match the required number density
of Al$_2$O$_3$ grains of the best-fit dust shell models up to roughly 
2 $\times$ the inner dust shell radius, or 4 $\times$ the photospheric
radius. However, we also showed that it is unlikely that Al$_2$O$_3$ 
grains alone can explain typical observed dust-to-gas ratios of the outflows
of Mira variables, owing to the relatively low abundance of aluminum. 
It is more likely that Al$_2$O$_3$ grains are present at 
close radii of $\sim$2--4 stellar radii and can serve as seed
particles for further dust formation at larger radii and lower
dust temperatures. In the case of group (I), these additional grains
need to preserve the spectral signature of Al$_2$O$_3$, in particular
the broad feature at 9--15\,$\mu$m, without adding features that
are not observed for these targets, such as the silicate features
at 9.8\,$\mu$m. We demonstrated that already the addition of 
small amounts of warm silicates to an Al$_2$O$_3$ shell has
a strong effect on the photometry and visibility spectra, and would
have been detected by our VLTI/MIDI observations.
Candidates for such additional grains metallic iron or 
(Fe,Mg)O (magnesiow\"{u}stite), 
\citep[e.g.,][]{Ferrarotti2006, McDonald2010,Posch2002}.
Also scattering iron-free magnesium-rich silicates
(Mg$_2$SiO$_4$, forsterite) may not have a significant
silicate signature in the mid-IR if distributed in certain geometries,
such as a geometrically thin shell \citep{Ireland2005}.
\citet{Sacuto2013} could explain VLTI/MIDI spectra of RT~Vir, which
have a similar shape as those of S~Ori and R~Cnc, by a combination
of forsterite grains and Al$_2$O$_3$ grains.
The presence of dust grains at small radii of $\sim$\,2 stellar radii and 
below have also been observationally confirmed by interferometric polarimetry
and been attributed to forsterite or Al$_2$O$_3$ grains
\citep{Ireland2005,Norris2012}. Dust grains at close  
radii of $\sim$\,2 stellar radii are co-located with SiO maser emission 
for all our sources, so that SiO nucleation and silicate dust formation
are unlikely at these radii.
Sources of group (I) are consistent with relatively low mass-loss rates,
so that the dust condensation sequence may freeze out for these sources
before the formation of silicates that display the typical silicate
features at 9.7\,$\mu$m and 18\,$\mu$m.

Sources of group (III), which can be described by a
mix of an Al$_2$O$_3$ shell and a silicate shell, are consistent 
to be associated with higher mass-loss rates and higher optical depths
of the dust shells. For these sources, the density at larger radii
may be sufficiently high so that the dust condensation sequence
described above may continue to the formation of amorphous silicates
such as olivine and pyroxene, which add the typical silicate emission
features to the spectra of these sources.

Sources of group (II), which can be described by a silicate dust
shell alone, may be of the same kind as sources of group (III), but
where the signature of Al$_2$O$_3$ grains is not detected. Alternatively,
these sources may be an example of direct silicate formation at larger
radii via SiO nucleation \citep{Gail2013} or heteromolecular condensation
\citep{Goumans2012}.

The results from the project will allow refinements and enhancements
of state-of-the-art dynamic model atmospheres and radiative transfer
modeling codes, as well as new models describing the mass-loss process
and the wind driving mechanism. Future observations aiming at
characterizing and constraining such new models would benefit from
obtaining additional concurrent spectrally resolved near-infrared
interferometry bringing stronger constraints on atmospheric molecular
layers located close to the photosphere.

\begin{acknowledgements}
  We would like to thank Prof. H.-P. Gail for his scientific 
  advice and his suggestions, and we would like to also thank the 
  referee for the comments that helped to improve our manuscript. 
  We used the AAVSO International Database along with the AFOEV and
  SIMBAD databases, operated at the CDS, France. We acknowledge star
  observers providing active data support in the observations of the
  variable stars. The authors would also like to acknowledge all who
  are involved in developing the publicly accessible MIDI data
  reduction software packages and tools EWS and MIA. 
  This work is based on service mode observations made with
  the MIDI instrument, which is operated by ESO. We would like to
  thank the operating team at the Paranal Observatory for their
  careful execution of the observations. Any findings or conclusions 
  expressed by DAB in this material
are those of the author and do not necessarily reflect the views
of the National Science Foundation.
\end{acknowledgements}

%\begin{thebibliography}{}
\bibliographystyle{aa}
\bibliography{KAROVICOVA}

\end{document}